\documentclass[12pt]{article}
\usepackage{latexsym}
\usepackage{epsf}
\usepackage{graphicx}

\def\be{\begin{equation}}
\def\ee{\end{equation}}
\def\bea{\begin{eqnarray}}
\def\eea{\end{eqnarray}}
\def\by{\left(\begin{array}}
\def\ey{\end{array}\right)}

\def\slash#1{\setbox0=\hbox{$#1$}#1\hskip-\wd0\dimen0=5pt\advance
       \dimen0 by-\ht0\advance\dimen0 by\dp0\lower0.5\dimen0\hbox
                to\wd0{\hss\sl/\/\hss}}

\def\ma0{\mbox{\boldmath $0$}}
\def\maq{\mbox{\boldmath $q$}}

\def\mak{\mbox{\boldmath $k$}}

\begin{document}

\begin{center}
\Large{$\rho - \omega$ splitting and mixing in nuclear matter}
\end{center}

\begin{center}
{\large S. Zschocke, B. K\"ampfer}
\end{center}

\begin{center}
\small{
Forschungszentrum Rossendorf, PF 510119, 01314 Dresden, Germany} 
\end{center}

\normalsize

\begin{abstract}

We investigate the splitting and mixing 
of $\rho$ and $\omega$ mesons in nuclear matter.
The calculations were performed on the basis of QCD sum rules and include 
all operators up to mass dimension-6 twist-4 and up to first order in the 
coupling constants. Special attention is devoted to 
the impact of the scalar 4-quark condensates on both effects. 
In nuclear matter the Landau damping governs the 
$\rho - \omega$ mass splitting while the scalar 4-quark condensates govern 
the strenght of individual mass shifts. 
A strong in-medium mass splitting causes the disappearance of the 
$\rho - \omega$ mixing.  

\end{abstract}
\section{Introduction}

The investigation of in-medium modifications of hadrons is currently a topic 
of wide interest. This is because the issue is related to chiral symmetry 
restoration as well as to a change of vacuum properties, and the phenomenon 
"mass of particles". Among the promising candidates for a search for changed 
hadron properties in an ambient strongly interacting medium are vector mesons. 
Due to their decay mode $V \rightarrow \gamma^{\star} \rightarrow {\rm e}^+ 
{\rm e}^-$ and the negligible interaction of the escaping 
${\rm e}^+ {\rm e}^-$ one can expect to probe directly the parent vector 
meson $V$. Indeed, strong evidences for changes of the $\rho$ meson are found 
in relativistic heavy-ion collisions, where a meson-rich hot medium is 
transiently created (cf. \cite{Shift5}). As the vector meson properties are 
coupled to various condensates \cite{Brown_Rho, lit11, lit12}, which change as 
a function of both the baryon density and the temperature, complementary  
investigations of their behavior via the ${\rm e}^+ {\rm e}^-$
decay channel in compressed nuclear matter is also required. 
Experimentally this will be done 
in a systematic way with the detector system HADES \cite{HADES}. The situation 
is quite challenging since various predictions differ in details. 

In the invariant mass region up to $1$ GeV there are various sources of 
${\rm e}^+ {\rm e}^-$ \cite{Cassing,TVN1}: Dalitz decays of many hadrons, 
bremsstrahlung, and the direct decays 
$V \rightarrow {\rm e}^+ {\rm e}^-$ mentioned above. 
One important channel for di-electron 
production is the reaction $\pi \pi \rightarrow \gamma^{\star} \rightarrow 
{\rm e}^+ {\rm e}^-$. 
This channel has been evaluated with increasing sophistication over the last 
years (cf. \cite{Shift5}).
The corresponding di-electron production rate 
$R = d N^{\rm ee} / d^4 x$ in a medium characterized by the baryon 
density $n$ and temperature $T$ is given by 
\cite{productionrate_1,productionrate_5}
\bea
\frac{d R}{d M_{\rm e e}} (M_{\rm e e}, n, T) 
= \frac{\sigma (M_{\rm ee}^2, n)}{(2 \pi)^4}
M_{\rm e e}^{4} \;T\;K_1 \left(\frac{M_{\rm e e}}{T}\right)\;
\left(1 - \frac{4\, m_{\pi}^2}{M_{\rm e e}^2}\right)\;.
\label{mixing_sumrule_65}
\eea
Here $K_1$ is a modified Bessel function, $M_{\rm e e}$ stands for the 
invariant mass of the di-electron pair, and $\sigma$ is the total cross 
section of the process $\pi^+ \pi^- \;\rightarrow \gamma^{\star} 
\rightarrow {\rm e}^+ {\rm e}^-$  
\bea
\sigma (q^2, n) = \frac{4}{3} \pi\frac{\alpha_{\rm em}^2}
{q^2}
\sqrt{1 - \frac{4 \, m_{\pi}^2}{q^2}} \; 
|{\rm F}_{\pi} (q^2, n)|^2\;,
\label{mixing_sumrule_70}
\eea
where ${\rm F}_{\pi} (q^2)$ is the pion formfactor and $q^2 = M_{\rm e e}^2$ 
is the momentum squared of the decaying virtual photon $\gamma^{\star}$. 

The $\rho - \omega$ mixing in vacuum has been discovered as a particular 
form of the pion formfactor $|F_{\pi} (q^2)|$ a few decades ago 
\cite{Augustin}. Since that time much work has been done 
aiming at a theoretical understanding of this mixing effect  
(for a review see \cite{Connell1}). Despite that there is still some debate 
concerning details of the $\rho - \omega$ mixing in vacuum 
\cite{Mixing15,Maltman,Leinweber} the experi\-mental pion formfactor in vacuum 
can well be reproduced by means of several theoretical approaches. 
However, up to now the mixing effect has not been studied systematically 
in the medium. One may argue that the $\rho - \omega$ mixing is a tiny 
effect in evaluating the di-electron emission rate of warm nuclear matter. 
Moreover, inspired by \cite{Mixing5}, it seems to be a generic effect of 
its own interest, which should be analyzed in a dense medium. 
This is one issue of the present paper. 
In addition, we are going to investigate the $\rho - \omega$ mass splitting 
and mixing effect simultaneously on the same footing, i.e. we use the 
same parameter set and the same approach for evaluating both effects. 

Refs.~\cite{Mixing5, zschocke1,zschocke2} showed that $\rho$ and $\omega$ 
mesons experience, within the 
QCD sum rule approach, quite a different in-medium behavior. Even in 
zero-width approximation a large $\rho - \omega$ mass splitting was found 
when neglecting terms in the operator product expansion (OPE) 
which differ for $\rho$ and $\omega$ mesons.
The individual mass shifts depend on the yet poorly known  
density dependence of the four-quark condensate, assuming the same effective 
four-quark condensate for both the $\rho$ and the $\omega$ mesons.
A further goal of the present paper is to include all terms in the OPE up 
to mass dimension-6 and twist-4 (up to order $\alpha_s$) and to study the 
importance of condensates which make $\rho$ and $\omega$ differ.
We extend our previous studies \cite{zschocke1,zschocke2} to consider 
here the yet unexplored effect of the density dependence of the four-quark 
condensate on the $\rho - \omega$ mixing in medium. 

Our paper is organized as follows. 
In section 2 we determine the mass shifts of $\rho$ and $\omega$ mesons. 
We spell out the basic steps of QCD sum rules in low-density 
approximation and list all terms of the operator product expansion (OPE) 
up to mass dimension-6 and twist-4. We then present a numerical evaluation 
of the QCD sum rules and show that terms in the OPE which make $\rho$ and 
$\omega$ differ are small at nuclear matter saturation density. The 
$\rho - \omega$ mass splitting is found to be determined by different 
Landau damping terms, while the individual mass shifts are governed by the 
density dependence of the four-quark condensate. The  
knowledge of the in-medium $\rho, \omega$ mass parameters is a prerequisite 
of a consistent treatment of the $\rho - \omega$ mixing studied   
in section 3.
We define the phenomenology of the mixing and explain how this effect is 
related to observables. Afterwards we specify the QCD sum rule for the 
$\rho - \omega$ mixing and present details of the evaluation. The summary 
can be found in section 4.

\section{$\rho - \omega$ mass splitting}

The masses of $\rho$ and $\omega$ mesons differ in vacuum by a small amount, 
$\Delta m = m_{\omega} - m_{\rho} = 11$ MeV. It was one success of the QCD sum 
rule method to explain this mass splitting in vacuum by differences in the 
OPE of $\rho$ and $\omega$ current$-$current correlators \cite{lit11}. 
Indeed, up to mass dimension-6 there is only one operator in vacuum, 
the so-called flavor mixing scalar operator, which 
differs in sign in the OPE of $\rho$ and $\omega$ correlators. 
In the following we will investigate the behavior of this splitting 
at finite density, where nonscalar condensates play also a role.

\subsection{QCD sum rule}

Within QCD sum rules the in--medium vector mesons $V=\rho, \omega$ are 
considered as resonances
in the current$-$current correlation function
\bea
\Pi_{\mu \nu}^{(V)} (q , n) = i \int d^4 x \;{\rm e}^{i q x}\;
\langle \Omega |\, {\rm T} \; {\rm J}_{\mu}^V (x)\; {\rm J}_{\nu}^V (0) 
\, | \Omega \rangle\;,
\label{eq_5}
\eea
where $q_{\mu}=(q_0, \maq)$ is the meson four momentum, T denotes the
time ordered product of the meson current operators
${\rm J}_{\mu}^{V} (x)$, and $|\Omega\rangle$ stands for a state of the 
nuclear medium.
In the following, we focus on the ground state of baryonic matter approximated
by a Fermi gas with nucleon density $n$ (in Ref. \cite{zschocke1} it was 
shown that 
temperature effects for $T\le 100$ MeV are of subleading order and may be 
neglected for our purposes). We first study isospin symmetric nuclear 
matter and extend later on our approach to asymmetric nuclear matter.
In terms of quark field operators, the vector meson currents are given by
\bea
{\rm J}_{\mu}^{V} = \frac{1}{2} (\overline{u} \gamma_{\mu}  u
\mp \overline{d} \gamma_{\mu} d)\;,
\label{vectorcurrent}
\eea
where the upper sign stands for the $\rho$ meson while the lower sign stands 
for 
the $\omega$ meson. We will keep this notation troughout the paper. 
Note that the interpolating currents eq.~(\ref{vectorcurrent}) are based on 
the same field operators $u,d$. Therefore, evaluating the r.h.s. of 
eq.~(\ref{eq_5}) 
will deliver the same condensates, however a few of them with 
different signs. To highlight this point we spell out all terms 
arising from eqs.~(\ref{eq_5}, \ref{vectorcurrent}) in the following.

We consider the nucleon and vector meson at rest,
i.e. $q_{\mu} = (q_0, \maq = 0)$ and $k_{\mu} = (M_N, \mak=0)$, which
implies the vector meson to be off shell while the nucleon is on shell.
Then the correlator (\ref{eq_5}) can be reduced to
$\frac{1}{3} \Pi_{\mu}^{\mu} (q^2, n) = \sum\limits_{V=\rho,\omega}
\Pi^{(V)} (q^2, n)$. In each of the vector meson channels the
correlator
$\Pi^{(V)}(q^2, n)$ satisfies the twice subtracted dispersion relation,
which can be written with $Q^2 \equiv -q^2 = -E^2$ as
\bea
\frac{\Pi^{(V)} (Q^2, n)}{Q^2} = \frac{\Pi^{(V)} (0,n)}{Q^2} - \Pi^{(V) '}
(0) + Q^2 \frac{1}{\pi}\,
\int\limits_0^{\infty} \; ds \frac{{\rm Im} \Pi^{(V)}(s, n)}{s^2 (s + Q^2)}
\label{eq_10}
\eea
with $\Pi^{(V)} (0,n) = \Pi^{(V)} (q^2=0, n)$ and $\Pi^{(V) '} (0)=
\frac{{\rm d} \Pi^{(V)} (q^2, n)}{{\rm d} q^2}|_{q^2=0}$
as subtraction constants.
We use $\Pi^{(\rho)} (0,n) = n/(4 M_N)$ and $\Pi^{(\omega)} (0,n) =
9\, n/(4 M_N)$ \cite{Mixing5,TVN2}, 
respectively, which are the Thomson limit of the $V N$ scattering process 
and correspond to Landau damping terms \cite{Landau5}.

As usual in QCD sum rules \cite{lit11,lit12}, for large values of $Q^2$
one can evaluate the nonlocal operator of 
eq.~(\ref{eq_5}) by OPE. We truncate the OPE 
beyond mass dimension-6 and twist-4 and include all terms up to 
the first order in $\alpha_s$ in the SU(2) flavor sector:  
\bea
\Pi^{(V)} (Q^2, n) &=& \Pi^{(V)}_{\rm scalar}
+ \Pi^{(V)}_{d=4,\tau=2}
+ \Pi^{(V)}_{d=6,\tau=2}
+ \Pi^{(V)}_{d=6,\tau=4}\; + \;.\;.\;.\;,
\label{eq_15}
\\
\nonumber\\
\Pi^{(V)}_{\rm scalar} &=&
- \frac{1}{8 \pi^2}
\left(1+\frac{\alpha_s}{\pi} \,C_F\,\frac{3}{4} \right) \;Q^2 \;
{\rm ln} \left(\frac{Q^2}{\mu^2}\right)
- \frac{3}{8 \pi^2} \left(m_u^2 + m_d^2 \right)
\label{ope_1}
\\
&& + \frac{1}{2} 
\left(1 + \frac{\alpha_s}{\pi} \, C_F \,
\frac{1}{4} \,\right) \frac{1}{Q^2} \; \langle \Omega|
(m_u \,\overline{u} u + m_d \,\overline{d} d)
|\Omega\rangle + \frac{1}{24} \frac{1}{Q^2}
\langle\Omega|\frac{\alpha_s}{\pi}
{\rm G}^2|\Omega\rangle
\label{ope_2}
\\
&& - \frac{1}{2} \pi \alpha_s \frac{1}{Q^4} 
\langle\Omega|\left(
\overline{u} \gamma_{\mu}\gamma_5\lambda^a u\;
\overline{u} \gamma^{\mu}\gamma_5\lambda^a u \;+\; 
\overline{d} \gamma_{\mu}\gamma_5\lambda^a d\;
\overline{d} \gamma^{\mu}\gamma_5\lambda^a d\right) |\Omega\rangle
\label{ope_4}
\\
&& \pm \pi \alpha_s \frac{1}{Q^4} \langle\Omega|
\left(\overline{u} \gamma_{\mu}\gamma_5\lambda^a u\;
\overline{d} \gamma^{\mu}\gamma_5\lambda^a d\right) |\Omega\rangle
\label{ope_5}
\\
&& - \frac{1}{9} \pi \alpha_s \frac{1}{Q^4} \langle\Omega|
\left(\overline{u} \gamma_{\mu}\lambda^a u\; 
\overline{u} \gamma^{\mu}\lambda^a u \;+\;
\overline{d} \gamma_{\mu}\lambda^a d  \;
\overline{d} \gamma^{\mu}\lambda^a d\right)
|\Omega\rangle 
\label{ope_7}
\\
&& - \frac{2}{9} \pi \alpha_s \frac{1}{Q^4} \langle\Omega|
\left(\overline{u} \gamma_{\mu}\lambda^a u \;
\overline{d} \gamma^{\mu}\lambda^a d\right)
|\Omega\rangle \;,
\label{ope_8}
\\
&& + g_s \frac{1}{12} \frac{1}{Q^6} \left( m_u^2 \langle \Omega | 
m_u\, \overline{u} \sigma_{\mu \nu}\,G^{\mu \nu}  u | \Omega \rangle + 
m_d^2 \; \langle \Omega | m_d \, \overline{d} \sigma_{\mu \nu}\,G^{\mu \nu} 
d | \Omega \rangle \right)\;,
\label{ope_new1}
\\
\Pi^{(V)}_{d=4,\tau=2} &=& 
\frac{1}{2}\, \frac{\alpha_s}{\pi} n_f \, \frac{1}{Q^4} q^{\mu} q^{\nu}
\langle\Omega| \hat {\rm S} \hat {\rm T} \left(
G_{\mu}^{\;\;\alpha} G_{\alpha \nu}
\right) |\Omega\rangle
\label{ope_10}
\\
&& - \left( \frac{2}{3} 
- \frac{\alpha_s}{\pi} \,C_F\, \frac{5}{18} \right) \; i \;
\frac{1}{Q^4} q^{\mu} q^{\nu}
\langle\Omega| \hat {\rm S} \hat {\rm T} \left(\overline{u}
\gamma_{\mu} D_{\nu} u +
\overline{d} \gamma_{\mu} D_{\nu} d \right) |\Omega\rangle\;,
\label{ope_9}
\\
\nonumber\\
\Pi^{(V)}_{d=6,\tau=2} &=&  
- \frac{41}{27} \frac{\alpha_s}{\pi} \, n_f 
\frac{1}{Q^{8}} q^{\mu} q^{\nu} q^{\lambda} q^{\sigma}
\langle\Omega| \hat {\rm S} \hat {\rm T} \left(
G_{\mu}^{\;\;\rho} D_{\nu} D_{\lambda} G_{\rho \sigma}
\right)|\Omega\rangle 
\label{ope_12}
\\
&& \hspace{-1.8cm} + \left( \frac{8}{3}  
+ \frac{\alpha_s}{\pi} C_F \frac{67}{30} \right) \;i\;
\frac{1}{Q^{8}} q^{\mu} q^{\nu} q^{\lambda} q^{\sigma}
\langle\Omega| \hat {\rm S} \hat {\rm T} \left(\overline{u}
\gamma_{\mu} D_{\nu} D_{\lambda} D_{\sigma} u +
\overline{d} \gamma_{\mu} D_{\nu} D_{\lambda} D_{\sigma} d \right)
|\Omega\rangle\;,
\label{ope_11}
\\
\nonumber\\
\Pi^{(V)}_{d=6,\tau=4} &=& \pm \frac{1}{3} \frac{1}{Q^{6}}
q^{\mu} q^{\nu} \langle\Omega| g_s^2 \, 
\hat {\rm S} \hat {\rm T}\left(
\overline{u}\gamma_{\mu} \gamma_5 \lambda^a u\;
\overline{d}\gamma_{\nu} \gamma_5 \lambda^a d \right)|\Omega\rangle
\label{ope_18}
\\
&& \hspace{-1.8cm} - \frac{1}{6} \frac{1}{Q^{6}}
q^{\mu} q^{\nu} \langle\Omega| g_s^2 \, \hat {\rm S} \hat {\rm T}
\left(
\overline{u}\gamma_{\mu} \gamma_5 \lambda^a u \;
\overline{u}\gamma_{\nu} \gamma_5 \lambda^a u \;+\; 
\overline{d}\gamma_{\mu} \gamma_5 \lambda^a d \;
\overline{d}\gamma_{\nu} \gamma_5 \lambda^a d \right) |\Omega\rangle
\label{ope_17}
\\
&&\hspace{-1.8cm} - \frac{1}{24} 
\frac{1}{Q^{6}} q^{\mu} q^{\nu} \langle\Omega| 
g_s^2 \,\hat {\rm S} \hat {\rm T}
\left( \overline{u}\gamma_{\mu} \lambda^a u 
\left(\overline{u}\gamma_{\nu} \lambda^a u \;+\; 
\overline{d}\gamma_{\nu} \lambda^a d 
\right) \right) | \Omega\rangle
\label{ope_14}
\\
&&\hspace{-1.8cm} - \frac{1}{24} 
\frac{1}{Q^{6}} q^{\mu} q^{\nu} \langle\Omega| 
g_s^2 \, \hat {\rm S} \hat {\rm T}
\left(\overline{d}\gamma_{\mu} \lambda^a d
\left(\overline{u}\gamma_{\nu} \lambda^a u \;+\;
\overline{d}\gamma_{\nu} \lambda^a d 
\right) \right) |\Omega\rangle	 
\label{ope_15}
\\
&&\hspace{-1.8cm} - \frac{5}{12} \frac{1}{Q^{6}} q^{\mu} q^{\nu}
 \langle\Omega| i g_s 
 \hat {\rm S} \hat {\rm T}
 \left(\overline{u} \left[D_{\mu} , \tilde{G}_{\nu \alpha}
 \right]_{+}
 \gamma^{\alpha} \gamma_5 u +
 \overline{d} \left[D_{\mu} , \tilde{G}_{\nu \alpha} \right]_{+}
 \gamma^{\alpha} \gamma_5 d \right) |\Omega\rangle
\label{ope_19}
\\
&& \hspace{-1.8cm} - \frac{7}{3} \frac{1}{Q^6} q^{\mu} q^{\nu} \langle\Omega| 
\hat {\rm S} \hat {\rm T} \left( m_u \;\overline{u} D_{\mu} D_{\nu} u 
+ m_d \; \overline{d} D_{\mu} D_{\nu} d \right) | \Omega \rangle \;,
\label{ope_20}
\eea
where $n_f = 3$ is the number of active flavors at a scale of 1 GeV, 
and $C_F = (n_c^2 -1)/(2 n_c) = 4/3$  with $n_c = 3$ as number of colors; 
$\sigma_{\mu \nu} = \frac{i}{2} \left[\gamma_{\mu} \;, \gamma_{\nu} \right]_{-}$. 
The strong couplings are related by $\alpha_s=g_s^2/(4\pi)$. 

The SU(3) color matrices are normalized as 
Tr $(\lambda^a\,\lambda^b)=2 \delta^{a b}$, the covariant derivative is defined
as $D_{\mu}=\partial_{\mu}+i g A_{\mu}^{a}\lambda^a/2$ and $G^2=G^a_{\mu \nu}
G^{a\; \mu \nu}$ where $G_{\mu \nu}^a$ is the gluon field strength tensor
($G^{\mu \nu} = G^{a\; \mu \nu} \lambda^a/2$).
The dual gluon field strength tensor is defined by $\tilde{G}_{\mu \nu}
= \epsilon_{\mu \nu \rho \sigma} G^{a \;\rho \sigma} \lambda^a/2$.

The OPE for scalar operators up to mass dimension-6 can be found in 
\cite{lit11}.
For the twist-2 condensates we have included all singlet 
operators with even parity up to order $\alpha_s$ 
(nucleon matrix elements of operators with odd parity vanish).  
Their Wilson coefficients can be deduced from \cite{Buras}. 
\footnote{Despite the fact that twist-2 non-singlet operators occur in the 
OPE for electromagnetic currents \cite{Buras}, they are absent in the OPE 
of eq.~(\ref{eq_5}). 
Twist-2 non$-$singlet condensates contribute, however,  
to $\rho-\omega$ mixing.}

The Wilson coefficients of the twist-4 operators in lines 
(\ref{ope_18}) ... (\ref{ope_19}) are given in 
\cite{productionrate_1}, and for the twist-4 operator in line (\ref{ope_20}) 
it can be deduced from \cite{twist_quark}, where it has been found that the 
term (\ref{ope_20}) has some relevance for twist-4 effects of nucleon 
structure functions. 
The Wilson coefficient of an additional dimension-6 twist-4 operator, 
$\hat {\rm S} \hat {\rm T} \overline{q} \left[ D_{\mu} ,
G_{\nu \alpha} \right]_{-} \gamma^{\alpha} q $ (for an estimate of this 
condensate see \cite{Kim_Lee}), vanishes \cite{twist_quark,twist_4}.

We emphasize that the only difference between $\rho$ and $\omega$
mesons in the truncated OPE consists in the terms in lines (\ref{ope_5}) and 
(\ref{ope_18}). As mentioned above 
the term in line (\ref{ope_5}) is responsible for the $\rho - \omega$ mass 
splitting in vacuum; the term in line (\ref{ope_18}) vanishes in vacuum.
It is now our goal to analyze the in-medium difference of $\rho$ and $\omega$ 
mesons stemming from the OPE side. 
Most terms in lines (\ref{ope_1}) ... (\ref{ope_20}) may be evaluated using 
standard techniques \cite{Leupold1}. 
However, what remains to be considered are the flavor-mixing condensates in 
the lines (\ref{ope_5}), (\ref{ope_8}), (\ref{ope_18}), the mixed 
quark-gluon condensate in line (\ref{ope_new1}), the pure gluonic condensates 
in the lines (\ref{ope_10}), (\ref{ope_12}) and the twist-4 condensate 
in line (\ref{ope_20}).
The QCD corrections to order $\alpha_s$ of the twist-2 
condensates in lines (\ref{ope_9}) and 
(\ref{ope_11}) have not been taken into account in previous analyses. 

The chiral condensate and scalar gluon condensate 
have been discussed in some detail in \cite{cond5}. Details for the scalar 
four-quark condensates in lines (\ref{ope_4}) ... (\ref{ope_8}) are given in 
Appendixes A and B, where also further notations are explained. 
The twist-2 quark condensates (lines (\ref{ope_9}) and (\ref{ope_11})) and the 
gluonic twist-2 condensates (lines (\ref{ope_10}) and (\ref{ope_12})) are 
explicitly given in Appendix C.
The twist-4 condensates (lines (\ref{ope_18}) ... (\ref{ope_20})) are listed 
in Appendix D. 

Performing a Borel transformation \cite{lit11}
of the dispersion relation eq.~(\ref{eq_10}) with appropriate
mass parameter $M^2$ and taking into account the
OPE (\ref{eq_15}) one gets the QCD sum rule 
\bea
\Pi^{(V)} (0,n) - \frac{1}{\pi} 
\int\limits_0^{\infty} d s \;\frac{{\rm Im} \Pi^{(V)}(s,n)}{s}\; 
{\rm e}^{-s/M^2} =
c_0\; M^2 + \sum\limits_{i=1}^{\infty} \frac{c_i}{(i-1)! M^{2 (i-1)}}\,.
\label{eq_25}
\eea
For nuclear matter we utilize the one-particle dilute gas approximation
in order to evaluate all relevant condensates in the nuclear medium, i.e.
\bea
\langle \Omega| \hat {\cal O} |\Omega\rangle &=&
\langle \hat {\cal O} \rangle_0
+ \frac{n}{2 M_N} \langle N(\mak)| \hat{\cal O}|N(\mak)\rangle\;,
\label{eq_40}
\eea
where the nucleon states are normalized by 
$\langle N(\mak)| N(\mak')\rangle = (2\pi)^3\; 2\, E_{k}\;
\delta(\mak-\mak')$ with $E_k = \sqrt{\mak^2+M_N^2}$ .
The scalar dimension-4 and dimension-5 condensates are given by 
\cite{condensates} 
\bea
m_u \;\langle \Omega| \overline{u} u |\Omega\rangle 
&=& m_u \langle \overline{u} u \rangle_0 + \frac{1}{2} \sigma_N^u \;n\;, 
\label{condD}
\\
\langle \Omega| \frac{\alpha_s}{\pi} G^2 |\Omega\rangle &=&
\langle \frac{\alpha_s}{\pi} G^2 \rangle_0 -
\frac{8}{9} M_N^0\,n\;,
\label{condE}
\\
\langle \Omega| g_s \overline{u} \sigma_{\mu \nu} G^{\mu \nu} u | \Omega 
\rangle &=& \lambda^2 \langle \overline{u} u \rangle_0 + \frac{1}{2} \lambda^2 
\frac{\sigma_N^u}{m_u} \;n\;, 
\label{condmixed}
\eea
where we have introduced the sigma term 
$\sigma_N^u = m_u \langle N(\mak)| \overline{u} u | N(\mak) \rangle / M_N$ 
and $\lambda^2 \simeq 1 \;{\rm GeV}^2$. 
The chiral $d$ quark 
condensate follows by replacing the $u$ quark by a $d$ quark. 
The nucleon sigma term \cite{condensates} is 
$2 \sigma_N = \sigma_N^u + \sigma_N^d$. 

Inserting the explicit expressions for all condensates (cf. Appendixes A ... D) 
one gets for the coefficients $c_{1,2,3}$ in eq.~(\ref{eq_25})
\bea
c_0 &=& \frac{1}{8 \pi^2} \left(1 + \frac{\alpha_s}{\pi} \, C_F \frac{3}{4}
\right)\;, 
\label{eq_46}\\
c_1 &=& - \frac{3}{8 \pi^2} (m_u^2 + m_d^2) \;,\\
c_2 &=& \frac{1}{2} \left(1 + \frac{\alpha_s}{\pi} \, C_F \frac{1}{4} \right) 
\left(m_u \langle\overline{u} u\rangle_0  
\;+\; m_d \langle\overline{d} d\rangle_0 + \sigma_N\,n
\right) \nonumber\\
&& + \frac{1}{24} \left[\langle\frac{\alpha_s}{\pi} G^2 \rangle_0
- \frac{8}{9} M_N^0 \;n\right]\nonumber\\
&& + \left(\frac{1}{4} - \frac{5}{48}  \frac{\alpha_s}{\pi} C_F \right) 
A_2^{(u+d)} \;M_N \;n 
- \frac{3}{16} \,n_f \,\frac{\alpha_s}{\pi} A_2^G\,M_N\,n\;,
\label{eq_47}\\
c_3 &=& - \frac{112}{81} \pi \;\alpha_s \;\kappa_0\;
\langle\overline{q} q\rangle_0^{2}
\left[1+\frac{\kappa_N}{\kappa_0}\frac{\sigma_N}{m_q
\langle \overline{q} q\rangle_0}\;n\right]
\nonumber\\
&& + \frac{(8 \pm 36)}{81} \frac{Q_0^2}{f_{\pi}^2} \frac{\alpha_s}{\pi} 
\langle \overline{q} q \rangle_0^2
\left[1 +  \frac{\sigma_N}{m_q} \frac{1}
{\langle \overline{q} q \rangle_0}  \;n\right]\nonumber\\ 
&& - \left(\frac{5}{12} + \frac{\alpha_s}{\pi}\, C_F\, \frac{67}{192} \right) 
A_4^{(u + d)} M_N^3\;n\; 
+ \;\frac{205}{864} \;\frac{\alpha_s}{\pi} \,n_f\, A_4^G \;M_N^3\;n \nonumber\\
&& + \frac{1}{4} M_N \,n\,\left(\frac{3}{8} K^2_u + 
\frac{3}{2} K^1_u - (1 \pm 1) K^1_{ud} + 
\frac{15}{16} K^g_u \right)\nonumber\\
&& - \frac{7}{144} \, \sigma_N \, M_N^2 \;n\;,
\label{eq_48}
\eea
where $A_n^{(u + d)} = A_n^u + A_n^d $ with $n=2,4$, 
$2 \langle \overline{q} q \rangle_0 = \langle \overline{u} u \rangle_0 + 
\langle \overline{d} d \rangle_0$, and $2 m_q = m_u + m_d$. 

\subsection{Evaluation}

We define a ratio of weighted moments  
\bea
m_V^2 (n,M^2,s_V) \equiv 
\frac{\int\limits_0^{s_V} ds \; {\rm Im} \Pi^{(V)} (s,n)
\;{\rm e}^{-s/M^2}}
{\int\limits_0^{s_V} ds \; {\rm Im} \Pi^{(V)} (s,n)\;s^{-1}\; 
{\rm e}^{-s/M^2}}  
\label{mass}
\eea
for which the desired sum rule follows by taking the ratio of  
eq.~(\ref{eq_25}) to its derivative with respect to $1/M^2$ as 
\bea
m_V^2 (n,M^2,s_V) = \frac{c_0\,M^2\,[1-\left(1+\frac{s_V}{M^2}\right) 
{\rm e}^{-s_V/M^2}]
-\frac{c_2}{M^2} - \frac{c_3}{M^4}}{c_0\,\left(1-{\rm e}^{-s_V/M^2}\right)
+ \frac{c_1}{M^2} + \frac{c_2}{M^4} + \frac{c_3}{2 M^6} - 
\frac{\Pi^{(V)} (0,n)}{M^2}}\;\;,
\label{eq_70}
\eea
where we have identified the highlying (continuum) contributions as 
$ - {\rm Im} \Pi^{(V)} (s \ge s_V,n)/s = \pi c_0$ ($s_V$ is the continuum 
threshold).
The meaning of the parameter $m_V^2$ as normalized first moment of the spectral 
function ${\rm Im} \Pi^{(V)}$ becomes immediately clear in zero-width 
approximation, $ - {\rm Im} \Pi^{(V)} (s\le s_V, n)= \pi F_V 
\delta(s - m^{\star \;2}_V)$,  
from where $m_V=m^{\star}_V$ follows. eqs.~(\ref{mass}, \ref{eq_70}) are the 
corresponding generalizations for the case of finite width, in the spirit of a 
resonance + continuum ansatz. 
The mass equation (\ref{eq_70}) is commonly used for describing $m_V^2$ 
in vacuum
\cite{lit11,Narison,zschocke4,litD1,massequation1,massequation2},
at finite temperature \cite{productionrate_1} and at finite density
\cite{lit12,Mixing5,zschocke1} and will be subject of our 
further considerations. 

The sum rule is reliable only in a Borel window $M^2_{\rm min} \le M^2 \le 
M^2_{\rm max}$. If $M^2$ is too small the expansion eq.~(\ref{eq_25})
breaks down. On the other side, if $M^2$ is too large the contribution
of perturbative QCD terms completely dominate the sum rule.
We adopt the following rules for determining the
Borel window \cite{Leupold1,Leupold2,Leupold3,eval10,eval15}:
The minimum Borel mass, $M^2_{\rm min}$, is determined such that the terms
of order $O (1/M^6)$ on the OPE side contribute not more than $10 \%$.
The maximum Borel mass, $M^2_{\rm max}$, is evaluated within 
zero-width approximation by requiring that the continuum part is not larger 
than the contribution of the resonance part, i.e.   
\bea
\frac{1}{8 \pi^2} \left(1+\frac{\alpha_s}{\pi}\right) M_{\rm max}^2 \; 
{\rm e}^{-s_V/M^2_{\rm max}}
\le \frac{F_V}{m_V^2} \;{\rm e}^{-m_V^2/M^2_{\rm max}}
\;.
\label{parameter_10}
\eea
The parameter $F_V$ can be evaluated by means of the QCD sum rule 
eq.~(\ref{eq_25}). 
The obtained results for vacuum, 
$F_{\rho} = 0.0110$ GeV$^4, F_{\omega} = 0.0117$ GeV$^4$, are in good 
agreement with the relations
$F_{\rho} = m_{\rho}^4/g^2_{\rho \gamma} = 0.0130$ GeV$^4$ 
and $F_{\omega} = 9\;m_{\omega}^4/g^2_{\omega \gamma} = 0.0138$ GeV$^4$, 
respectively, which follow from the Vector Meson Dominance (VMD) 
\cite{lit11,Connell1,Leupold2}. 

The threshold $s_V$ is determined by maximum flatness of 
$m_V (n,M^2,s_V)$ as a function of $M^2$. 
These requirements give a coupled system of equations 
for the five unknowns $M^2_{\rm min}$, $M^2_{\rm max}$, $F_V$, $s_V$, $m_V$.
The final parameters $\overline{F}_V$ and $\overline{m}_V$ 
are averaged to get Borel mass independent quantities. 
For any parameter $P$ this average is defined by 
\bea
\overline{P} = \frac{1}{M^2_{\rm max} - M^2_{\rm min}}
\int\limits_{M^2_{\rm min}}^{M^2_{\rm max}} P (M^2) \; d M^2 \;.
\label{parameter_15}
\eea
In the following we will skip the average sign. 


\subsection{Results}

For considering the mass parameter splitting effect there is no 
need to distinguish between isospin symmetric and isospin asymmetric 
nuclear matter since all operators in the OPE eq.~(\ref{eq_15}) are 
isospin symmetric operators. Accordingly, in the following we 
study isospin symmetric nuclear matter. 
   
Twist-4 condensates have been estimated in \cite{DIS} where 
data of lepton-nucleon forward scattering amplitude has been used 
to fix the parameters $K_u^1$, $K_u^2$, $K_u^g$ and $K_{u d}^1$ in 
eq. (\ref{eq_48}).
The corresponding system of equations is under-determined and therefore 
various sets for these parameters can be obtained. 
We have investigated all six sets from \cite{DIS} for these parameters and find 
only very small changes of the results. In Fig.~\ref{fig: fig1} we show 
the results obtained with the parameter set given in Appendix D. 
Since in \cite{zschocke1,zschocke2,zschocke4,zschocke3} a strong effect of 
the density dependence of the four-quark condensate was found we show here 
results for various possibilities, parameterized by $\kappa_N$ introduced 
in Appendix A, eq.~(\ref{A_20}).
The mass parameter of the $\rho$ meson decreases with increasing density for 
all $\kappa_N$, while the $\omega$ meson mass parameter decreases only for 
sufficiently large $\kappa_N$. Other QCD sum rule analyses 
\cite{lit12,eval10,zschocke1} have obtained also a decreasing $\rho$ mass 
parameter.
An increase of the $\omega$ meson mass parameter has been found in 
\cite{Mixing5,zschocke1}, where the correct Landau damping term was 
implemented. 

\begin{figure}[!h]
\includegraphics[scale=0.3]{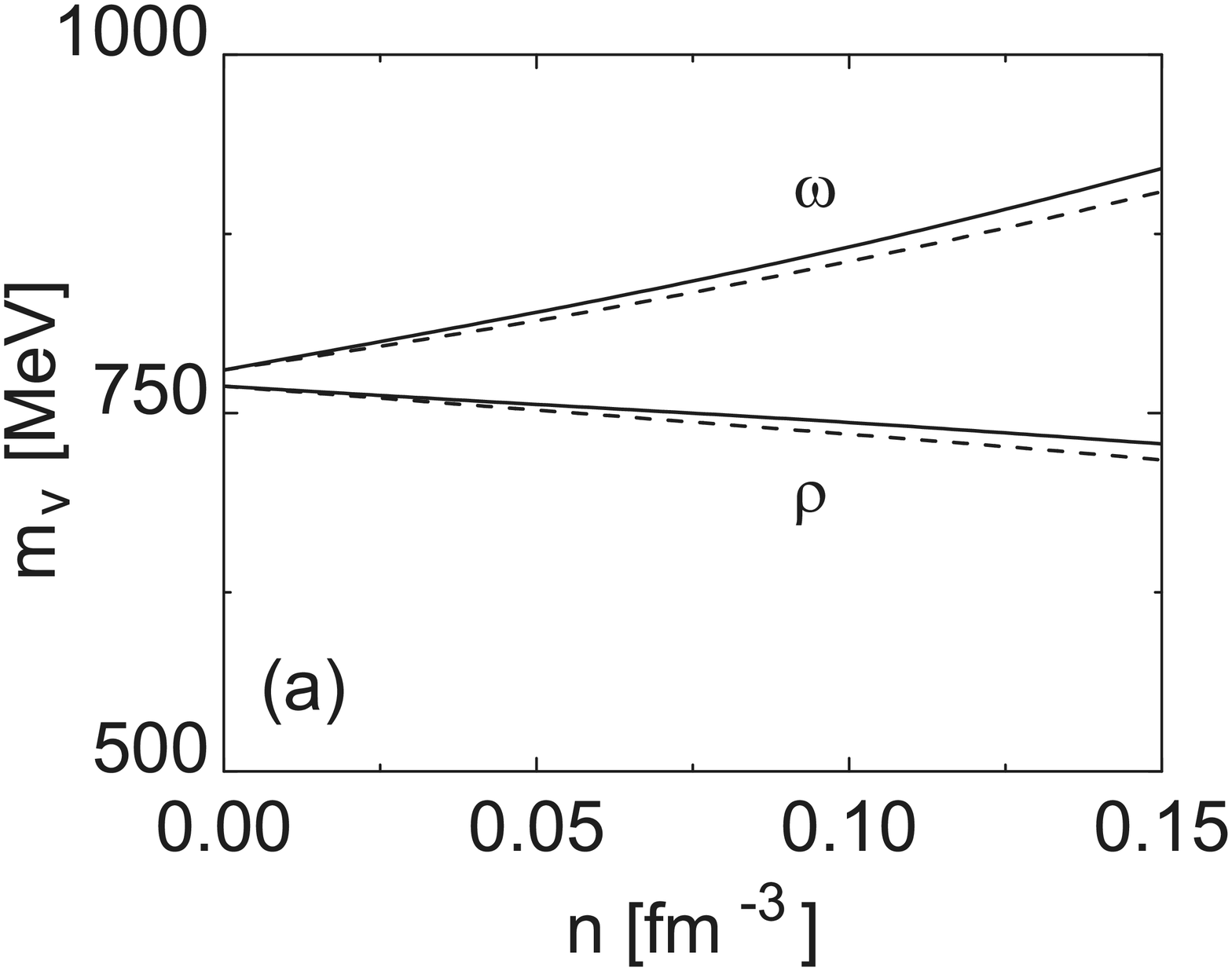}
\includegraphics[scale=0.3]{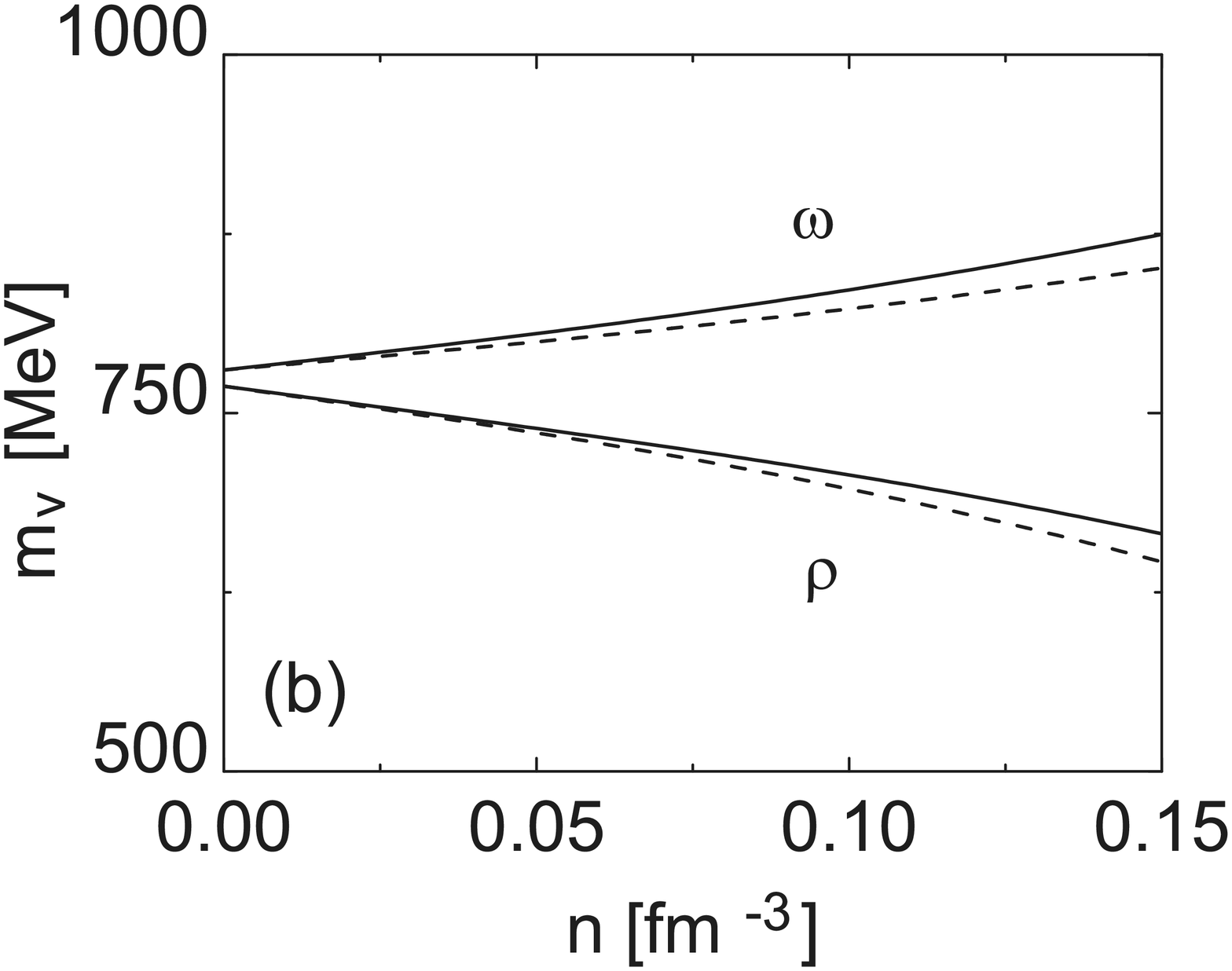}
\includegraphics[scale=0.3]{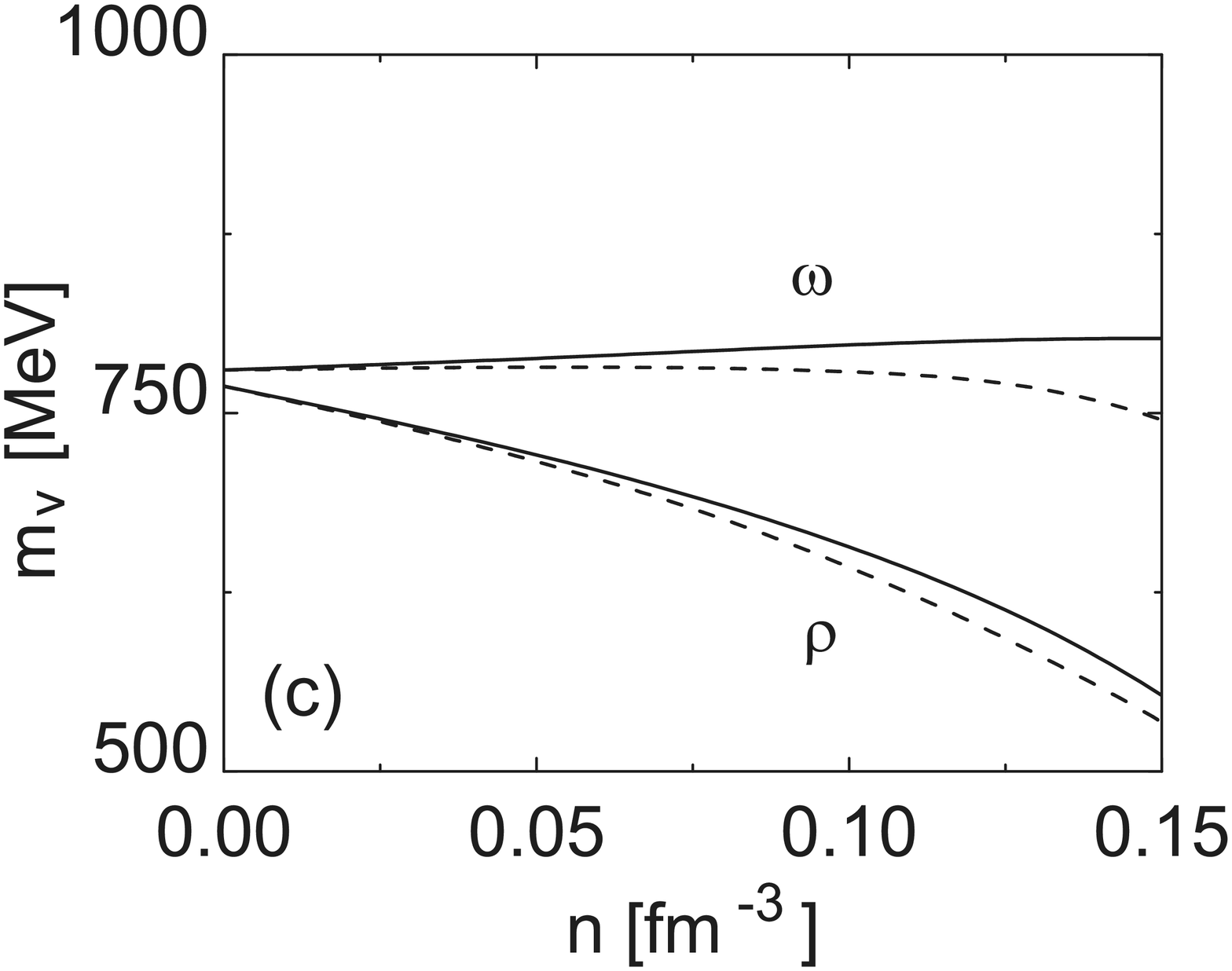}
\includegraphics[scale=0.3]{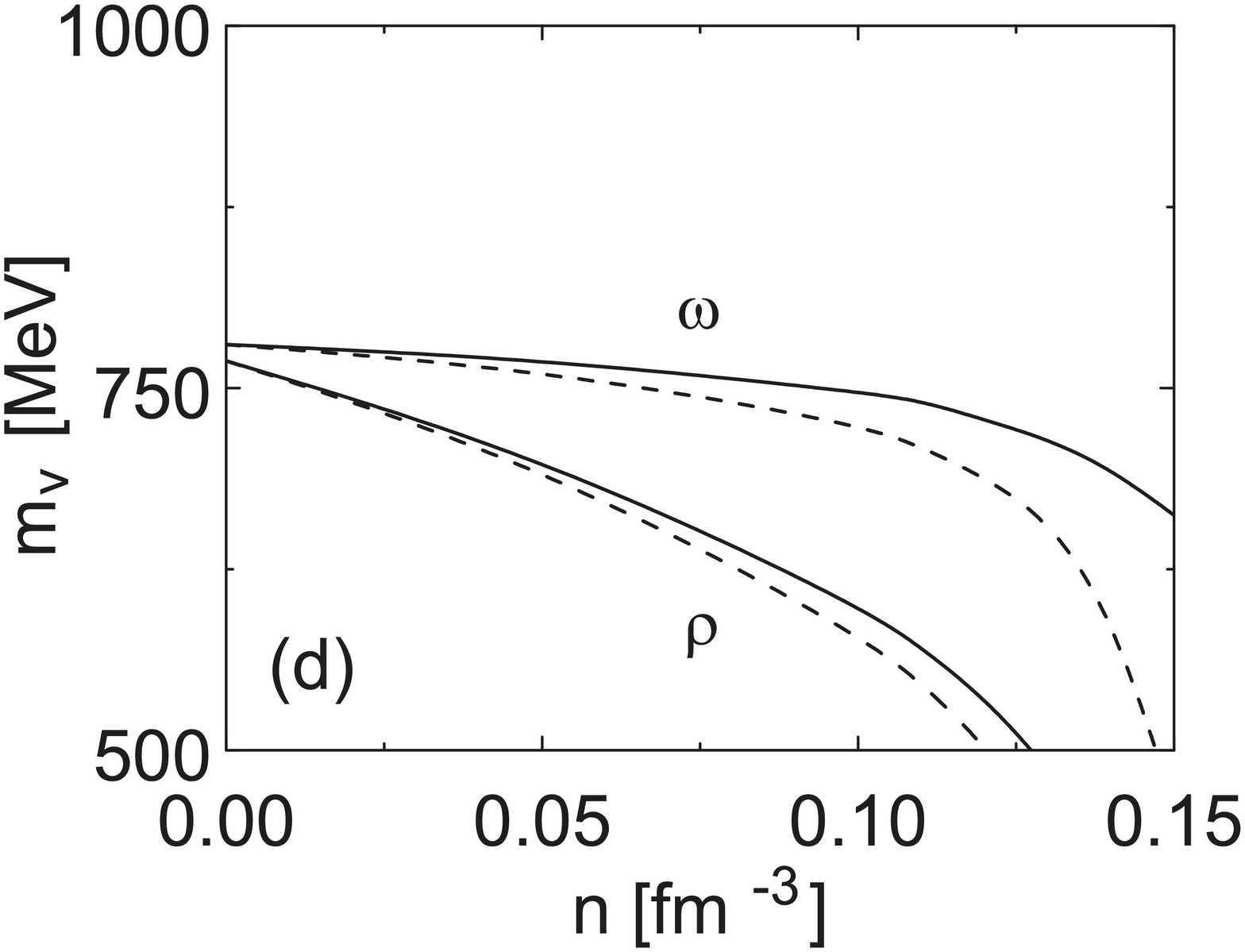}
\caption{Mass parameter $m_V$ of $\omega$ meson (upper curves) and $\rho$ meson
(lower curves) as a function of the density for various values of the 
parameter $\kappa_N$ ((a): $\kappa_N=1$, (b): $\kappa_N=2$, (c): $\kappa_N=3$, 
(d): $\kappa_N=4$). The solid curves are for the full set of terms in 
eqs.~(\ref{eq_46} - \ref{eq_48}),  
while for the dotted curves the twist-4 condensates are discarded, i.e. 
$K^{1,2,g}_{u,d,ud} = 0$.}
\label{fig: fig1}
\end{figure}


The flavor mixing scalar operators (i.e. $M^{u d}_{A,V}$, see Appendix B),
while responsible for the mass splitting in vacuum, play only a minor  
rule in matter. That means, discarding the terms $\sim Q_0^2$ in
eq.~(\ref{eq_48}) yields curves which are nearly identical with those 
represented in Fig.~\ref{fig: fig1}. 
The poorly known scalar four-quark condensate governs the strength
of individual mass shifts, while the strong mass splitting in matter 
originates mainly from the Landau damping terms $\Pi^{(V)} (0,n)$, 
which differ by a factor $9$ for $\rho$
and $\omega$ \cite{Mixing5}.
The outcome of our study is that terms in the OPE, which cause a 
difference of $\rho$ and $\omega$ mesons, are small in matter since the 
mass splitting is mainly determined by the Landau damping terms. 

\newpage

\section{$\rho - \omega$ mixing}

First, we briefly describe the mixing scenario considered in the 
following. We follow the arguments 
given in \cite{Connell1}. The mixing can be accomplished by
\bea
\left(\begin{array}[c]{l}
\displaystyle
\rho\\
\displaystyle
\omega
\end{array}
\right) =
\left(\begin{array}[c]{l}
\displaystyle
1 \quad - \epsilon\\
\displaystyle
\epsilon \quad\;\;\; 1
\end{array}
\right)
\left(\begin{array}[c]{l}
\displaystyle
\rho_I\\
\displaystyle
\omega_I
\end{array}
\right) \;,
\label{mixing_10}
\eea
where the subscript $I$ denotes isospin-pure states, and $\epsilon$ is the
mixing parameter.
The mixing formula (\ref{mixing_10}) is quite general.  
Extending the mixed propagator approach described in \cite{Connell1} 
to the case of finite density one can obtain the following relation between 
the complex mixing parameter $\epsilon$ and the nondiagonal 
selfenergy $\delta_{\rho\,\omega} (q^2,n)$ via (cf. \cite{Connell1} for vacuum, 
cf. \cite{Mixing5} for matter)
\bea
\epsilon (n) = \frac{\delta_{\rho\,\omega} (q^2, n)}{m_{\omega}^2 (n) -
m_{\rho}^2 (n) 
+ i\; {\rm Im} \Sigma_{\omega}(q^2,n) - i\; {\rm Im} \Sigma_{\rho} (q^2,n)}\;.
\label{mixing_20}
\eea
The nondiagonal selfenergy $\delta_{\rho\,\omega} (q^2,n)$, and therefore also 
the mixing parameter $\epsilon$, is directly related to the pion form factor, 
given by 
\bea 
F_{\pi} (q^2,n) &=& 1 - \frac{q^2}{g_{\rho \gamma}}\frac{g_{\rho \pi \pi}}
{q^2 - m_{\rho}^2 (n) - i\, {\rm Im} \Sigma_{\rho} (q^2,n)} 
\nonumber\\
&& - \frac{q^2}{g_{\omega \gamma}}
\frac{1}
{q^2 - m_{\omega}^2 (n) - i \,{\rm Im} \Sigma_{\omega} (q^2, n)}
\delta_{\rho\,\omega} (q^2, n)
\frac{g_{\rho \pi \pi}}
{q^2 - m_{\rho}^2 - i \,{\rm Im} \Sigma_{\rho} (q^2, n)}\;.
\label{mixing_30}
\eea
Since the main contribution of the second line stems from the region 
$q^2 \sim m_{\rho}^2, 
m_{\omega}^2$ one usually approximates the 
nondiagonal selfenergy $\delta_{\rho \omega} (q^2, n)$ in the pion formfactor 
by its on-shell value at $q^2 = \overline{m}^2 = 
0.5\, (m_{\rho}^2 + m_{\omega}^2)$.

The nondiagonal selfenergy consists of an electromagnetic part and a hadronic 
part, 

$\delta_{\rho\,\omega} (\overline{m}^2, n) =
\delta_{\rho\,\omega}^{\rm EM} (\overline{m}^2, n) +
\delta_{\rho\,\omega}^{\rm H} (\overline{m}^2, n)$.
Both contributions can consistently be isolated in theoretical as well
as experimental analyses.
The electromagnetic part comes from the process $\rho \rightarrow 
\gamma^{\star} \rightarrow \omega$ and can be evaluated analytically 
\cite{self_feynman}. 
In the following we are going to
investigate the density dependence of $\delta_{\rho\,\omega} 
(\overline{m}^2, n)$. 

\subsection{QCD sum rule}

The basic object for the $\rho - \omega$ mixing in matter is the mixed 
correlator
\bea
\Pi_{\mu \nu}^{\rho\;\omega} (q , n) = i \int d^4 x \;{\rm e}^{i q x}
\langle {\rm T} \; {\rm J}_{\mu}^{\rho} (x)\; {\rm J}_{\nu}^{\omega} (0)
\rangle_n \;, 
\label{mixing_sumrule_5}
\eea
with the isotriplet and isosinglet currents from eq.~(\ref{vectorcurrent}).
It is straightforward to recognize that
\bea
\Pi^{\rho\;\omega} (q, n) \equiv \frac{1}{3} g^{\mu \nu}\; 
\Pi_{\mu \nu}^{\rho\;\omega} (q, n) = \Pi^{\rm u}(q, n) - \Pi^{\rm d}(q, n) 
\label{mixing_sumrule_6}
\eea
with 
\bea
\Pi^{\rm q}(q, n) = \frac{i}{12}\, 
\int d^4 x \;{\rm e}^{i q x} \langle {\rm T} \; 
\overline{q}(x)\,\gamma_{\mu}\, q(x)\; 
\overline{q}(0)\,\gamma^{\mu}\,q(0) \rangle_n\;.
\label{mixing_sumrule_7}
\eea
This scalar function satisfies the twice subtracted dispersion relation 
\bea
\frac{\Pi^{\rho\;\omega} (q, n)}{Q^2} = \frac{\Pi^{\rho\;\omega} (0, n)}{Q^2} - 
\Pi^{\rho\;\omega\;'} (0, n)  + Q^2\;\frac{1}{\pi}\int\limits_0^{\infty} d s \;
\frac{{\rm Im} \Pi^{\rho\;\omega} (s,n)}{s^2 (s+Q^2)}\;.
\label{mixing_sumrule_10}
\eea
The subtraction constant $\Pi^{\rho\;\omega} (0, n)$ vanishes in vacuum 
\cite{Connell1} as  
well as in case of symmetric nuclear matter \cite{Mixing5}. 
For asymmetric nuclear matter, $\Pi^{\rho\;\omega} (0, n) = 
- 3 \;\alpha_{n p} \;n /(4 M_N)$
with $\alpha_{n p} = (n_n-n_p)/n$ \cite{Mixing5}, where $n_n$ and $n_p$ are 
the neutron and proton densities, respectively, and $n = n_n + n_p$. 
 
The other subtraction constant $\Pi^{\rho\;\omega\;'} (0, n) = 
\frac{d \Pi^{\rho\;\omega} (q^2, n)}{d q^2}|_{q^2=0}$ does not 
contribute to the sum rule after a Borel transformation. 
It is convenient \cite{Mixing10} to subtract the 
pure electromagnetic contribution 
$\rho \rightarrow \gamma^{\star} \rightarrow \omega$ from
hadronic and OPE sides of the dispersion relation (\ref{mixing_sumrule_10}).
In doing so we arrive at 
a new function, denoted by $\tilde{\Pi}^{\rho\;\omega}$, 
which satisfies the same dispersion relation  
eq.~(\ref{mixing_sumrule_10}).

For large values of $Q^2$ one evaluates
the l.h.s. of eq.~(\ref{mixing_sumrule_10}) by the OPE. 
Due to large cancellations of the pure QCD terms
according to eq.~(\ref{mixing_sumrule_6}) 
one has now to include also the electromagnetic contributions 
to the OPE in contrast to eq.~(\ref{eq_15}), where the electromagnetic 
terms are neglegible compared to the QCD terms. 
Accordingly, up to mass dimension-6 twist-4, and up to first order in  
$\alpha_s$ and $\alpha_{\rm em}$ the OPE is given by 
(for vacuum cf. \cite{Mixing15,Mixing10}, for matter cf. \cite{Mixing5})
\bea
\tilde{\Pi}^{\rho\;\omega} (Q^2) &=& \tilde{\Pi}^{\rho\;\omega}_{\rm scalar}
+ \tilde{\Pi}^{\rho\;\omega}_{d=4,\tau=2}
+ \tilde{\Pi}^{\rho\;\omega}_{d=6,\tau=2}
+ \tilde{\Pi}^{\rho\;\omega}_{d=6,\tau=4}\; + \; .\;.\;. ,
\label{mixing_sumrule_15}
\\
\nonumber\\
\tilde{\Pi}^{\rho\;\omega}_{\rm scalar} &=&
- \frac{1}{64 \pi^3} Q^2 \;\alpha_{\rm em}\; 
{\rm ln} \left(\frac{Q^2}{\mu^2}\right)
- \frac{3}{8 \pi^2} \left(m_u^2 - m_d^2 \right)
\label{mixing_sumrule_A}
\\
&& + \frac{1}{2} \left(1 + \frac{\alpha_s}{\pi} \,C_F\, \frac{1}{4}\right) 
\frac{1}{Q^2} \langle \Omega|
(m_u\, \overline{u} u - m_d \,\overline{d} d)
|\Omega\rangle 
\label{mixing_sumrule_B}
\\
&& + \frac{1}{72} \frac{\alpha_{\rm em}}{\pi}
\frac{1}{Q^2} \langle \Omega|
(4 \; m_u\, \overline{u} u - m_d \,\overline{d} d) |\Omega\rangle
\label{mixing_new2}
\\
&& - \frac{1}{2} \pi \alpha_s \frac{1}{Q^4} \left(\langle\Omega|
\overline{u} \gamma_{\mu}\gamma_5\lambda^a u\;
\overline{u} \gamma^{\mu}\gamma_5\lambda^a u|\Omega\rangle
- \langle\Omega|
\overline{d} \gamma_{\mu}\gamma_5\lambda^a d\;
\overline{d} \gamma^{\mu}\gamma_5\lambda^a d|\Omega\rangle
\right)
\label{mixing_sumrule_C}
\\
&& - \frac{1}{9} \pi \alpha_s \frac{1}{Q^4} \left(\langle\Omega|
\overline{u} \gamma_{\mu}\lambda^a u\;
\overline{u} \gamma^{\mu}\lambda^a u|\Omega\rangle
- \langle\Omega|
\overline{d} \gamma_{\mu}\lambda^a d\;
\overline{d} \gamma^{\mu}\lambda^a d|\Omega\rangle
\right)
\label{mixing_sumrule_D}
\\
&& - \frac{2}{9} \pi \alpha_{\rm em} \frac{1}{Q^4} 
\left( 4 \langle \Omega|
\overline{u} \gamma_{\mu}\gamma_5 u\;
\overline{u} \gamma^{\mu}\gamma_5 u|\Omega\rangle
- \langle\Omega|
\overline{d} \gamma_{\mu}\gamma_5 d\;
\overline{d} \gamma^{\mu}\gamma_5 d |\Omega\rangle \right)
\label{mixing_sumrule_E}
\\
&& - \frac{4}{81} \pi \alpha_{\rm em} \frac{1}{Q^4} \left( 
4 \langle\Omega| \overline{u} \gamma_{\mu} u\; 
\overline{u} \gamma^{\mu} u |\Omega\rangle
- \langle\Omega| \overline{d} \gamma_{\mu} d\; 
\overline{d} \gamma^{\mu} d |\Omega\rangle
\right)
\label{mixing_sumrule_F}
\\
&& + g_s \frac{1}{12} \frac{1}{Q^6} \left(
m_u^2 \;\langle \Omega | m_u \;\overline{u} \sigma_{\mu \nu} G^{\mu \nu} 
u | \Omega \rangle 
- m_d^2 \;\langle \Omega | m_d \; \overline{d} \sigma_{\mu \nu} G^{\mu \nu}
d | \Omega \rangle\right)
\label{mixing_new1}
\\
&& + e \frac{1}{12} \frac{1}{Q^6} \left( 
\frac{2}{3} m_u^2 \langle \Omega | 
m_u \; \overline{u} \sigma_{\mu \nu} F^{\mu \nu} u | \Omega \rangle  
+ \frac{1}{3} m_d^2 \langle \Omega | 
m_d \; \overline{d} \sigma_{\mu \nu} F^{\mu \nu} d | \Omega \rangle
\right)\;,
\label{mixing_new3}
\\
\tilde{\Pi}^{\rho\;\omega}_{d=4,\tau=2} &=& - \left( 
\frac{2}{3} - \frac{\alpha_s}{\pi} \,C_F\, \frac{5}{18} \right)  
\frac{i}{Q^4} q^{\mu} q^{\nu}
\langle\Omega| \hat {\rm S} \hat {\rm T} \left(\overline{u}
\gamma_{\mu} D_{\nu} u -
\overline{d} \gamma_{\mu} D_{\nu} d \right) |\Omega\rangle
\label{mixing_sumrule_G}
\\
&& + \frac{\alpha_{\rm em}}{\pi} \frac{5}{162} 
\frac{i}{Q^4} q^{\mu} q^{\nu}
\langle\Omega| \hat {\rm S} \hat {\rm T} \left( 4 \overline{u}
\gamma_{\mu} D_{\nu} u -
\overline{d} \gamma_{\mu} D_{\nu} d \right) |\Omega\rangle\;,
\label{mixing_new4}
\\
\nonumber\\
\tilde{\Pi}^{\rho\;\omega}_{d=6,\tau=2} &=& \hspace{-0.2cm} \left(\frac{8}{3} 
\!+\! \frac{\alpha_s}{\pi} \,C_F\, \frac{67}{30} \right)
\frac{i}{Q^{8}} q^{\mu} q^{\nu} q^{\lambda} q^{\sigma}
\langle\Omega| \hat {\rm S} \hat {\rm T} \left(\overline{u}
\gamma_{\mu} D_{\nu} D_{\lambda} D_{\sigma} u -
\overline{d} \gamma_{\mu} D_{\nu} D_{\lambda} D_{\sigma} d \right)
|\Omega\rangle\;\;
\label{mixing_sumrule_H}
\\
&& + \frac{\alpha_{\rm em}}{\pi} \frac{67}{270} 
\frac{i}{Q^{8}} q^{\mu} q^{\nu} q^{\lambda} q^{\sigma}
\langle\Omega| \hat {\rm S} \hat {\rm T} \left( 4 \;\overline{u}
\gamma_{\mu} D_{\nu} D_{\lambda} D_{\sigma} u -
\overline{d} \gamma_{\mu} D_{\nu} D_{\lambda} D_{\sigma} d \right)
|\Omega\rangle\;,
\label{mixing_new6}
\\
\nonumber\\
\tilde{\Pi}^{\rho\;\omega}_{d=6,\tau=4} &=&
 - \frac{1}{24}
\frac{1}{Q^{6}} q^{\mu} q^{\nu} \langle\Omega|
g_s^2 \,\hat {\rm S} \hat {\rm T}
\left( \overline{u}\gamma_{\mu} \lambda^a u
\overline{u}\gamma_{\nu} \lambda^a u \;-\;
\overline{d}\gamma_{\mu} \lambda^a d \overline{d}\gamma_{\nu} \lambda^a d
\right) | \Omega\rangle
\label{mixing_sumrule_I}
\\
&&\hspace{-1.8cm} - \frac{1}{6} \frac{1}{Q^{6}}
q^{\mu} q^{\nu} \langle\Omega| g_s^2 \, \hat {\rm S} \hat {\rm T}
\left(
\overline{u}\gamma_{\mu} \gamma_5 \lambda^a u\;
\overline{u}\gamma_{\nu} \gamma_5 \lambda^a u \;-\;
\overline{d}\gamma_{\mu} \gamma_5 \lambda^a d\;
\overline{d}\gamma_{\nu} \gamma_5 \lambda^a d \right) |\Omega\rangle
\label{mixing_sumrule_L}
\\
&&\hspace{-1.8cm} - \frac{5}{12} \frac{1}{Q^{6}} q^{\mu} q^{\nu}
\langle\Omega| i g_s\;
\hat {\rm S} \hat {\rm T}
\left(\overline{u} \left[D_{\mu} , \tilde{G}_{\nu \alpha}
\right]_{+}
\gamma^{\alpha} \gamma_5 u -
\overline{d} \left[D_{\mu} , \tilde{G}_{\nu \alpha} \right]_{+}
\gamma^{\alpha} \gamma_5 d \right) |\Omega\rangle
\label{mixing_sumrule_M}
\\
\nonumber\\
&& \hspace{-1.8cm} - \frac{7}{3} \frac{1}{Q^6} q^{\mu} q^{\nu} \langle\Omega|
\hat {\rm S} \hat {\rm T}
\left(m_u\; 
\overline{u} D_{\mu} D_{\nu} u - m_d \;\overline{d} D_{\mu}
D_{\nu} d \right) |\Omega\rangle
\label{mixing_sumrule_N}
\\
&& \hspace{-1.8cm} - \frac{1}{54}
\frac{1}{Q^{6}} q^{\mu} q^{\nu} \langle\Omega|
g_e^2 \,\hat {\rm S} \hat {\rm T}
\left( 4 \; \overline{u}\gamma_{\mu} u
\overline{u}\gamma_{\nu} u \;-\;
\overline{d}\gamma_{\mu} d 
\overline{d}\gamma_{\nu} d \right) | \Omega\rangle
\label{mixing_sumrule_Q}
\\
&&\hspace{-1.8cm} - \frac{2}{27} \frac{1}{Q^{6}}
q^{\mu} q^{\nu} \langle\Omega| g_e^2 \, \hat {\rm S} \hat {\rm T}
\left(
4 \; \overline{u}\gamma_{\mu} \gamma_5 u\;
\overline{u}\gamma_{\nu} \gamma_5  u \;-\;
\overline{d}\gamma_{\mu} \gamma_5 d\;
\overline{d}\gamma_{\nu} \gamma_5 d \right) |\Omega\rangle
\label{mixing_sumrule_S}
\\
&&\hspace{-1.8cm} - \frac{5}{12} \frac{1}{Q^{6}} q^{\mu} q^{\nu}
\langle\Omega| i \;e\;
\hat {\rm S} \hat {\rm T}
\left(\frac{2}{3} \;
\overline{u} \left[D^{\rm em}_{\mu} , \tilde{F}_{\nu \alpha}
\right]_{+}
\gamma^{\alpha} \gamma_5 u + \frac{1}{3}\;
\overline{d} \left[D^{\rm em}_{\mu} , \tilde{F}_{\nu \alpha} \right]_{+}
\gamma^{\alpha} \gamma_5 d \right) |\Omega\rangle\;.
\label{mixing_sumrule_T}
\eea
$\alpha_{\rm em} = e^2/(4 \pi)$ is the electromagnetic fine structure constant,
$F_{\nu \alpha}$ stands for the 
electromagnetic field strength tensor, and the dual electromagnetic field 
strength 
tensor is defined by $\tilde{F}_{\mu \nu} = \epsilon_{\mu \nu \rho \sigma} 
F^{\rho \sigma} $. The covariant derivative of QED is defined
as $D^{\rm em}_{\mu}=\partial_{\mu} + i e A_{\mu}$.
The QED contributions may be deduced from the QCD terms by the 
replacements $\lambda^a/2 \rightarrow 1$ (which implies $C_F \rightarrow 1$) 
and $g_s \rightarrow e_q$ ($e_q$ is the electric charge of quark $q$), 
respectively. Not all of the condensates given above have been taken into 
account in previous evaluations:
the terms in the lines (\ref{mixing_new2},  
\ref{mixing_new1}, \ref{mixing_new3}), the QCD corrections 
in the lines (\ref{mixing_sumrule_B}, \ref{mixing_sumrule_G}, 
\ref{mixing_sumrule_H}), the QED corrections given in the lines 
(\ref{mixing_new4}) and (\ref{mixing_new6}), and all twist-4 contributions 
in lines (\ref{mixing_sumrule_I} - \ref{mixing_sumrule_T}) have not been  
considered yet.

The isospin breaking of the scalar $u$ and $d$ quark condensates 
is usually parameterized by 
\bea
\gamma + 1 = \frac{\langle 0| \overline{d} d |0\rangle}
{\langle 0| \overline{u} u |0\rangle} \simeq 
\frac{\langle N| \overline{d} d |N\rangle}
{\langle N| \overline{u} u |N\rangle} \simeq
\frac{\langle \Omega| \overline{d} d |\Omega\rangle}
{\langle \Omega| \overline{u} u |\Omega\rangle} \;,
\label{mixing_sumrule_21}
\eea
where we have generalized the corresponding relation for vacuum
\cite{Mixing5,Mixing15,Mixing10} to the case of nuclear matter.

The four-quark condensates are given in Appendix A. The twist-2 
quark condensates are listed in Appendix C and the corresponding 
parameters can be found in Appendix E. 
The twist-4 condensates, listed here for the sake of completeness, 
are neglected in our analysis since they are strongly suppressed in the 
chosen Borel window. 

Performing a Borel transformation of eq.~(\ref{mixing_sumrule_10}) leads to 
\bea
\Pi^{\rho\, \omega} (0,n) - \frac{1}{\pi}
\int\limits_0^{\infty} d s \;\frac{{\rm Im} \tilde{\Pi}^{\rho \omega} 
(s,n)}{s}\;
{\rm e}^{-s/M^2} =
d_0\; M^2 + \sum\limits_{i=1}^{\infty} \frac{d_i}{(i-1)! M^{2 (i-1)}}\,.
\label{mixing_sumrule_25}
\eea
The coefficients $d_{1,2,3}$ in linear density approximation and neglecting 
all twist-4 condensates are given by
\bea
d_0 &=& \frac{1}{64 \pi^3} \alpha_{\rm em} \;, 
\label{mixing_sumrule_27}\\
d_1 &=& - \frac{3}{8 \pi^2} (m_u^2 - m_d^2) \;,
\label{mixing_sumrule_28}\\
d_2 &=& \left[ \frac{1}{2} \left(1 + \frac{\alpha_s}{\pi} 
\,C_F\,\frac{1}{4} 
\right) 
(m_u - m_d) +
\frac{1}{72} \frac{\alpha_{\rm em}}{\pi} \left(4 m_u - m_d \right) \right]
\left( \langle \overline{q} q \rangle_0 + \frac{\sigma_N}{2 m_q} n \right)
\nonumber\\
&&- \frac{\alpha_{\rm em}}{\pi} \frac{5}{288} M_N \;n\; \left(A_2^{u, p} 
+ A_2^{d, p} \right) \; + d_2^{\rm AS}\;,
\label{mixing_sumrule_29}\\ 
d_3 &=& \frac{14}{81} \pi \kappa_0 
\left(8 \gamma \alpha_s - \alpha_{\rm em} \right)
\langle \overline{q} q 
\rangle_0^2 \left(1 + \frac{\kappa_N}{\kappa_0} n \frac{\sigma_N}
{m_q \langle \overline{q} q \rangle_0}\right)\nonumber\\
&&  - \frac{\alpha_{\rm em}}{\pi} \frac{67}{1152} 
M_N^3 \;n \left(A_4^{u, p} + A_4^{d, p} \right) \; + d_3^{\rm AS}\;,
\label{mixing_sumrule_30}
\eea
where further terms proportional to $m_q \, \gamma, \gamma^2$ and 
$\gamma \;\alpha_{\rm em}$ 
have been neglected.  
The terms $d_l^{\rm AS}\; (l=2,3)$ are proportional to $\alpha_{n p}$ and 
account for isospin asymmetric matter. Their impact on   
mixing will be considered separately in subsection 3.3.3. 

Finally, we specify the hadronic side of the QCD sum rule 
eq.~(\ref{mixing_sumrule_25}) (cf. \cite{lit11, Mixing15, Maltman} for vacuum,  
\cite{Mixing5} for matter), where the $\phi$ meson has been implemented 
in accordance with \cite{Maltman}
\bea
- \frac{1}{\pi} \frac{{\rm Im} \tilde{\Pi}^{\rho\;\omega}(s,n)}{s} &=& 
\frac{1}{4} \left[f_{\rho} \; \delta(s-m_{\rho}^2) - 
f_{\omega} \; \delta(s-m_{\omega}^2) 
+ f_{\phi} \; \delta(s-m_{\phi}^2) \right]\nonumber\\
&& + \frac{1}{4} \left[f_{\rho'} \; \delta(s-m_{\rho'}^2) -  
f_{\omega'} \; \delta(s-m_{\omega'}^2)\right]  
 + \frac{\alpha_{\rm em}}
{64 \pi^3}\;\Theta(s-s_V)\;.
\label{mixing_sumrule_35}
\eea
The necessity for including the higher resonances $\rho'$ and $\omega'$ is 
discussed below.

We mention that the term $f_{\phi}$ is allowed since the $\phi$ meson is not 
a pure $\overline{s} s$ state but mixed with the $\omega$ meson.  
Even more, it has been found in \cite{Maltman} that
the $\phi$ meson gives a significant contribution in vacuum  
due to large cancellations between $f_{\rho}$ and $f_{\omega}$.  
Accordingly, we drop the assumption of ideal mixing
and take into account such a term. 
\footnote{In finite width QCD sum rule, which is necessary 
when considering the momentum dependence of mixing
$\delta_{\rho\,\omega} (q^2)$, the $\phi$ contribution is negligible 
\cite{Leinweber}.
We mention also that there is no need to use unphysical values for 
$m_{\rho'}$ and $m_{\omega'}$, as has been pointed out in \cite{Maltman}.} 

The five parameters $f_{\rho}, f_{\rho'}, f_{\omega},  f_{\omega'}$ and  
$f_{\phi}$ have to be evaluated selfconsistently within the QCD sum rule 
approach. 
What we still need is a connection between these new parameters and the 
parameter 
$\delta_{\rho\;\omega} (\overline{m}^2)$ which enters physical observables 
like the pion formfactor in eq.~(\ref{mixing_30}).
Such a relationship can be obtained by means of VMD \cite{Connell1}   
\bea
J^{\rho}_{\mu} (x) = \frac{m_{\rho}^2}{g_{\rho\, \gamma}} 
\varphi_{\mu}^{\rho} (x)\;,\quad\quad\;\quad\quad
J^{\omega}_{\mu} (x) = 3\;\frac{m_{\omega}^2}{g_{\omega\, \gamma}} 
\varphi_{\mu}^{\omega} (x)\;,
\label{eq_VMD}
\eea
where $\varphi_{\mu}^V (x)$ is the field operator of the respective vector 
meson $V = \rho, \omega$.
If one inserts these relations into the correlator eq.~(\ref{mixing_sumrule_5})
one gets an expression which relates $\Pi^{\rho\;\omega}_{\mu \nu}$
with the mixed propagator (keeping in mind the zero-width approximation 
at all stages).   
Another expression for $\Pi^{\rho\;\omega}_{\mu \nu}$ can be 
obtained by inserting eq.~(\ref{mixing_sumrule_35}) into the dispersion 
relation eq.~(\ref{mixing_sumrule_10}). Equating both expressions leads 
to the searched relation
\bea
\delta^{\rm H}_{\rho \;\omega} (\overline{m}^2) = 
- \left(f_{\rho} + f_{\omega}\right)\; \frac{1}{24}\; 
g_{\rho\,\gamma} g_{\omega\,\gamma} \; \frac{\Delta m^2}{\overline{m}^2} \;,
\label{mixing_sumrule_40}
\eea
which is valid to order ${\cal O} (\Delta m^4 / \overline{m}^4)$ being 
a fairly well approximation even when taking into account the 
strong mass splitting between $\rho$ and $\omega$ mesons.
We mention that for evaluating the momentum dependence of  
$\delta^{\rm H}_{\rho\,\omega} (q^2)$ the applicability of VMD has 
been debated in \cite{Maltman} due to the impact of the $\phi$ meson. 
On the other side, the reliablility of VMD for a momentum dependence 
of $\delta^{\rm H}_{\rho\,\omega} (q^2)$ has been confirmed in 
\cite{Leinweber}, where the finite width of vector mesons is taken 
into account. 
Anyhow, in zero-width approximation VMD, which leads to 
eq.~(\ref{mixing_sumrule_40}), is applicable as long as one  
restricts oneself to the on-shell value of this quantity, i.e. to 
$\delta^{\rm H}_{\rho\,\omega} (\overline{m}^2)$.

\subsection{Evaluation}

Since the most relevant parameter $\delta^{\rm H}_{\rho \;\omega}$ enters the 
approach via the combination $\zeta \sim f_{\rho} + f_{\omega}$ it is 
convinient to rewrite the sum rule eq.~(\ref{mixing_sumrule_25}) as 
\bea
\frac{1}{4} \zeta \frac{\overline{m}^2}{M^2} \left(
\frac{\overline{m}^2}{M^2} - \beta\right)\;{\rm e}^{-\overline{m}^2/M^2} 
+ \frac{1}{4} \zeta' \frac{\overline{m'}^2}{M^2} \left(
\frac{\overline{m'}^2}{M^2} - \beta'\right)\;{\rm e}^{-\overline{m'}^2/M^2} 
\; + \frac{1}{4} \frac{1}{M^2} f_{\phi}\; {\rm e}^{-m_{\phi}^2/M^2}
\nonumber\\
+\frac{\Pi^{\rho\,\omega} (0, n)}{M^2}\;+\; 
\frac{\alpha_{\rm em}}{64 \; \pi^3} {\rm e}^{-s_V/M^2} 
= d_0 + \sum\limits_{i=1}^{\infty} \frac{d_i}{(i-1)! M^{2 i}}\;,
\label{mixing_sumrule_45}
\eea
where we have introduced \cite{Maltman} 
\bea
\zeta &=& \frac{\Delta m^2}{\overline{m}^4}
\left(\frac{f_{\rho} + f_{\omega}}{2}\right)\;,
\quad\quad \quad\quad 
\zeta' = \frac{\Delta m'^2}{\overline{m'}^4}
\left(\frac{f_{\rho'} + f_{\omega'}}{2}\right)\;,
\nonumber\\
\beta &=& 2 \;\frac{f_{\omega} - f_{\rho}}{f_{\rho} + f_{\omega}}
\frac{\overline{m}^2}{\Delta m^2}\;,\quad\quad\quad \quad\quad
\beta' = 2 \; \frac{f_{\omega'} - f_{\rho'}}{f_{\rho'} + f_{\omega'}}
\frac{\overline{m'}^2}{\Delta m'^2}
\label{mixing_sumrule_50}
\eea
with $2 \overline{m}^2 = (m_{\rho}^2 + m_{\omega}^2)$, 
$2 \overline{m'}^2 = (m_{\rho'}^2 + m_{\omega'}^2)$, 
$\Delta m^2 = m_{\omega}^2 - m_{\rho}^2$ and 
$\Delta m'^2 = m_{\omega'}^2 - m_{\rho'}^2$, respectively.
We stress that eq.~(\ref{mixing_sumrule_45}) is valid to order 
${\cal O} (\Delta m^4 /M^4)$. Despite the observed strong mass splitting 
found in the previous section, eq.~(\ref{mixing_sumrule_45}) is a good  
approximation: The terms of order ${\cal O} (\Delta m^4 /M^4)$ would give less 
than 10 percent correction to terms of order ${\cal O} (\Delta m^2 /M^2)$, 
even at such a small Borel mass like $M \approx 1$ GeV.  

The residues in the hadronic model (\ref{mixing_sumrule_35}) can be 
expressed by the new variables (\ref{mixing_sumrule_50}) to give
\bea
f_{\rho} = \left(\frac{\overline{m}^2}{\Delta m^2} - \frac{\beta}{2} 
\right) \zeta\;
\overline{m}^2 \;,
\quad\quad \quad \quad
f_{\rho'} = \left(\frac{\overline{m'}^2}{\Delta {m'}^2} - \frac{\beta'}{2} 
\right) \zeta'\;\overline{m'}^2\;,
\\
f_{\omega} = \left(\frac{\overline{m}^2}{\Delta m^2} + \frac{\beta}{2} 
\right) \zeta
\;\overline{m}^2\;,
\quad\quad \quad\quad
f_{\omega'} = \left(\frac{\overline{m'}^2}{\Delta {m'}^2} 
+ \frac{\beta'}{2} \right) \zeta'\;\overline{m'}^2\;.
\label{mixing_sumrule_52}
\eea
Finally we give an expression for the mixing parameter $\epsilon$
in zero-width approximation (i.e., Im $\Sigma_{\rho,\omega}$ = 0)
which can be deduced from (\ref{mixing_sumrule_40}) and (\ref{mixing_20})
\bea
\epsilon = - \frac{\overline{m}^2}{\Delta m^2} \frac{g_{\rho \gamma} 
g_{\omega \gamma}}{12}\;\zeta\;.
\label{epsilon}
\eea

We need five equations for the five unknowns 
$\zeta,\zeta',\beta,\beta',f_{\phi}$. 
One could perform a Taylor expansion 
of eq.~(\ref{mixing_sumrule_25}) ending up with an equation system for these  
five parameters. This is the frame work of Finite Energy Sum Rules (FESR).
Instead, here we use a combined FESR and Borel analysis, 
following the approach described 
in \cite{Mixing15,Maltman} which we extend to finite density. 
Accordingly, the first equation comes from a local duality relation 
\cite{Mixing15}
which results into 
\bea
4\,\Pi^{\rho\,\omega} (0, n) \;-\beta\;\zeta\;\overline{m}^2 - \beta'\;\zeta'\;\overline{m'}^2 =
4\;d_0\;s_V + 4\;d_1 - f_{\phi} 
\label{phi_10}
\eea
and agrees with the first equation of the FESR approach \cite{Mixing15,Maltman}.
This equation makes clear why the higher resonances 
$\rho'$ and $\omega'$ have to be taken into account:  
Without these higher resonances one would get either $\beta\approx 0$ or 
$\zeta \approx 0$ which would be in contradiction with experimental findings. 
The second equation is just the sum rule eq.~(\ref{mixing_sumrule_25}). 
Two equations are obtained by the first and second derivatives  
with respect to $1/M^2$ of eq.~(\ref{mixing_sumrule_25}), cf. \cite{Mixing15} 
(due to the high Borel mass and the small contribution of 
the threshold term the second derivative sum rule is applicable, in contrast 
to the mass splitting, investigated in the previous section, where a second 
derivative sum rule becomes unstable \cite{eval10}).

For evaluating $f_{\phi}$ we still need a fifth equation.
In \cite{Maltman} a third derivative of sum rule has been used which could
cause instabilities due to the truncation of OPE \cite{eval10,eval15}.
To avoid such instabilities the individual
contributions of $\rho'$ and $\omega'$ have been approximated by an effective 
strenght $f_{\rho'\,\omega'}$ at the averaged mass of $m_{\rho'\,\omega'}$ 
in \cite{Leinweber}. However, at finite density $\beta$ is density dependent. 
Therefore, in order to improve this approximation we apply the second FESR for 
the parameter $\beta$, i.e., 
\bea
(1 + \beta)\;\zeta\;\overline{m}^4 + (1 + \beta')\;\zeta'\;\overline{m'}^4
-f_{\phi} \; m_{\phi}^2 = - 2\; d_0\;s_V^2 + 4\;d_2\;.
\eea
The resulting system of equations has to be solved selfconsistently giving the 
five unknowns as function of the Borel mass,
$\zeta(M^2), \zeta'(M^2), \beta(M^2), \beta'(M^2)$ and $f_{\phi} (M^2)$. 

\begin{figure}[!h]
\includegraphics[scale=0.2]{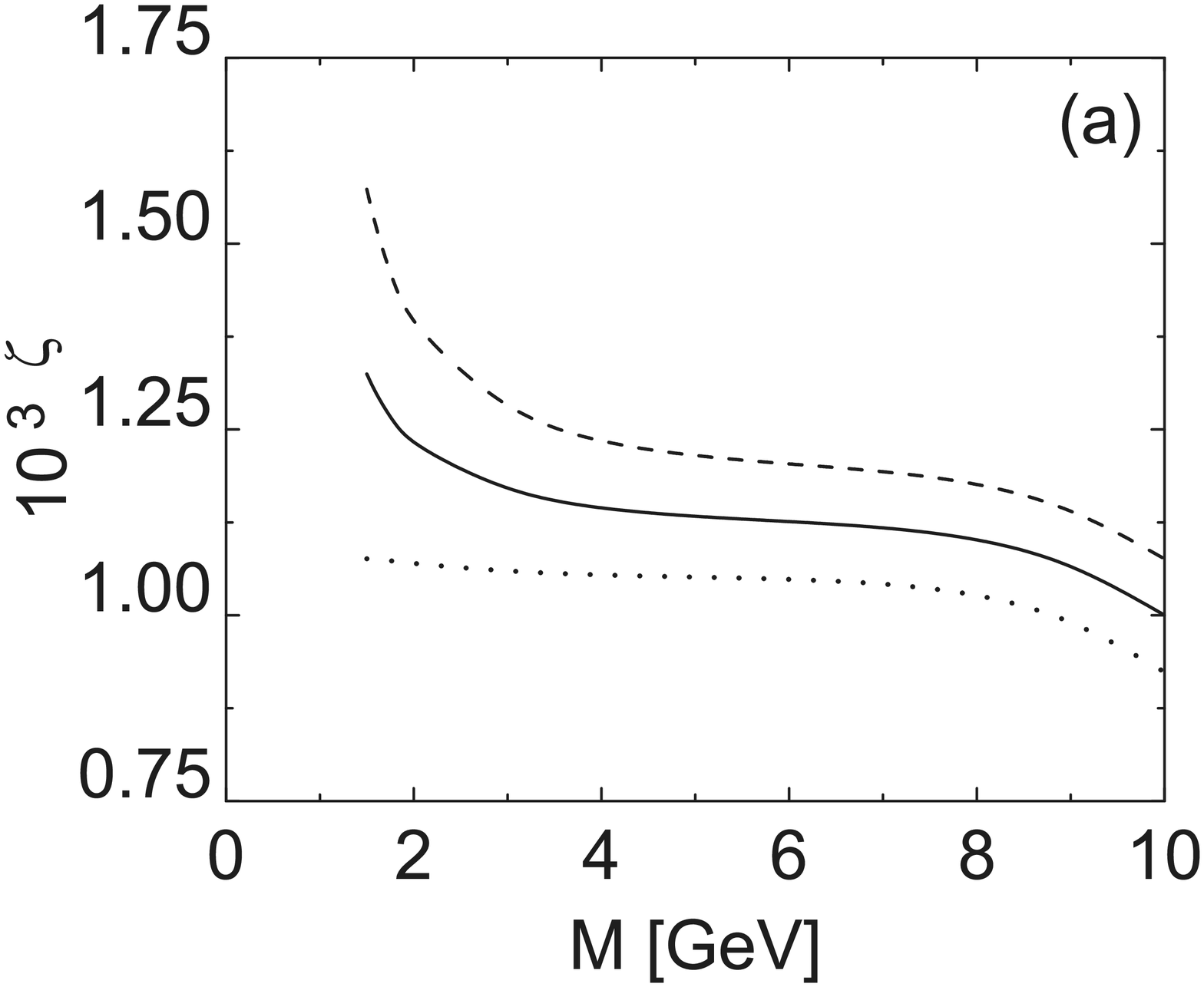}
\hspace{-0.5cm}
\includegraphics[scale=0.2]{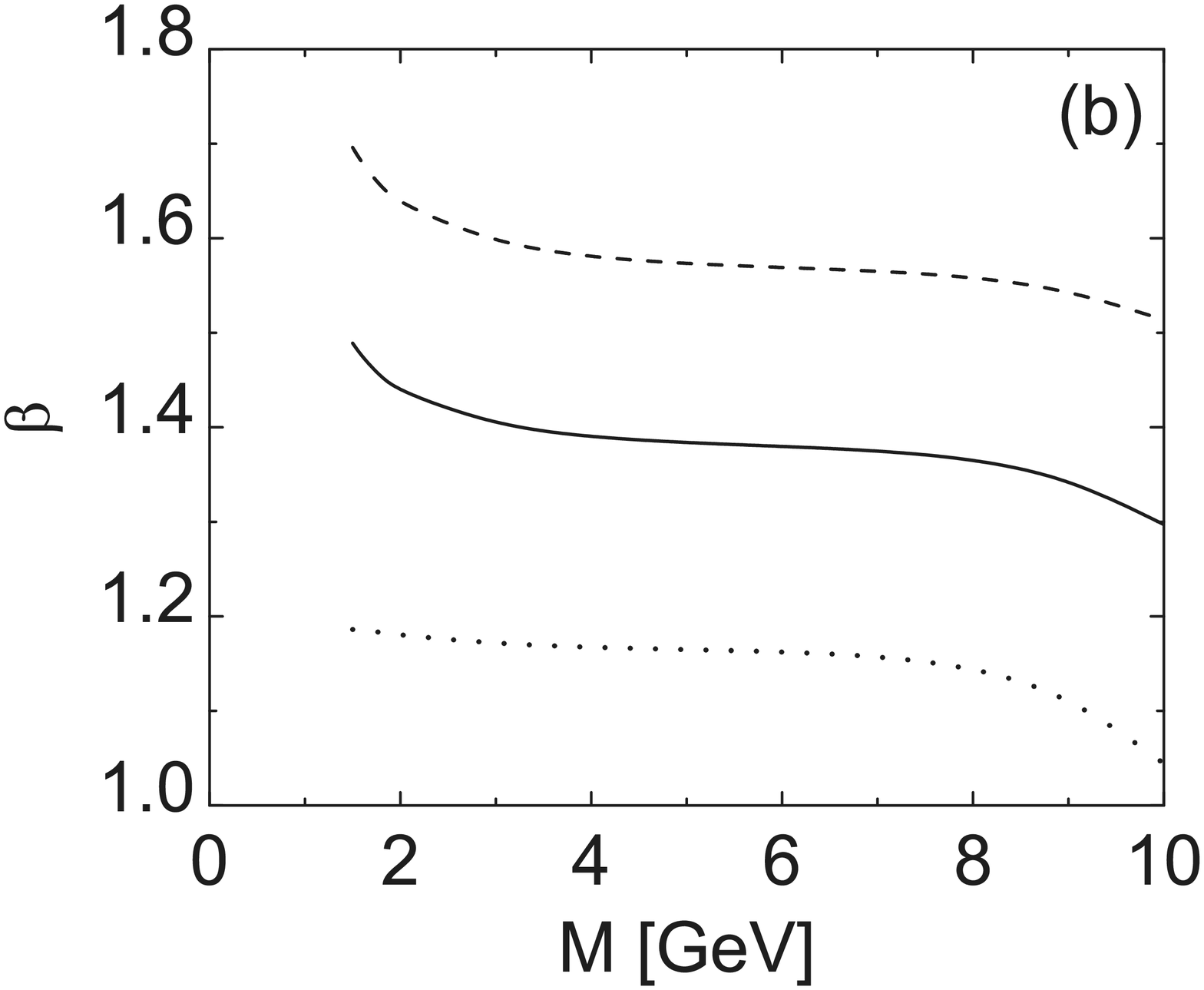}
\hspace{-0.5cm}
\includegraphics[scale=0.2]{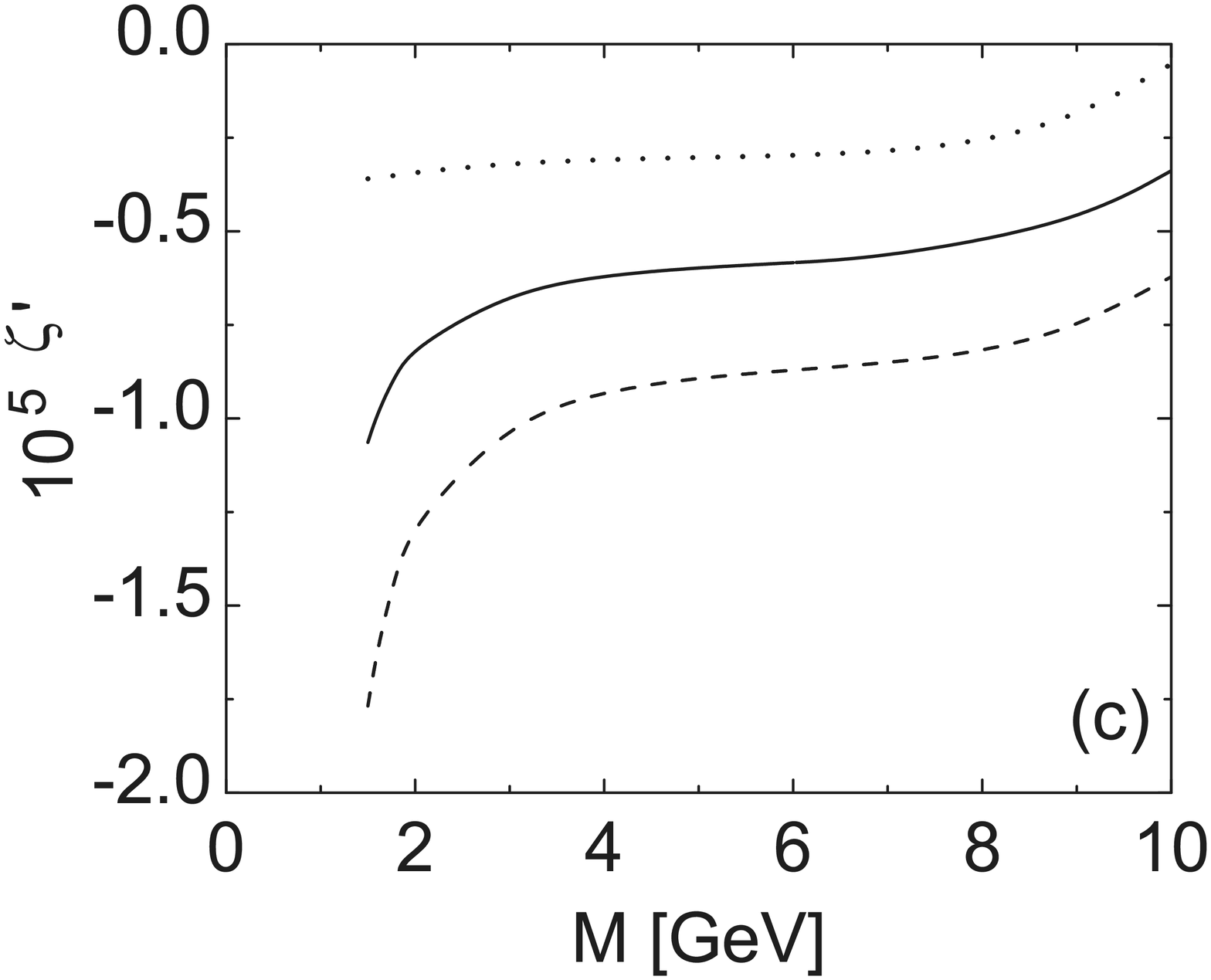}
\hspace{-0.5cm}
\includegraphics[scale=0.2]{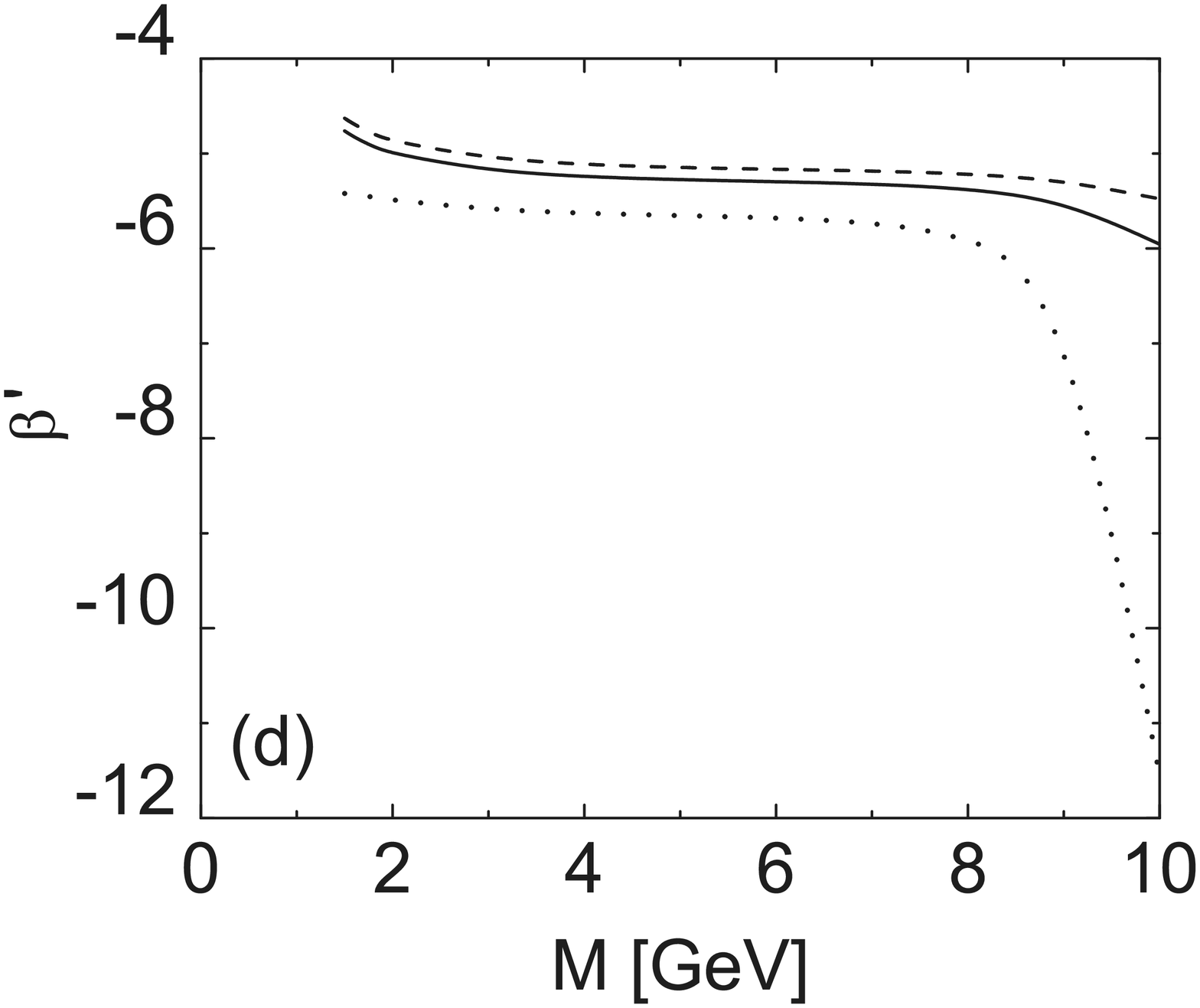}
\hspace{-0.5cm}
\includegraphics[scale=0.2]{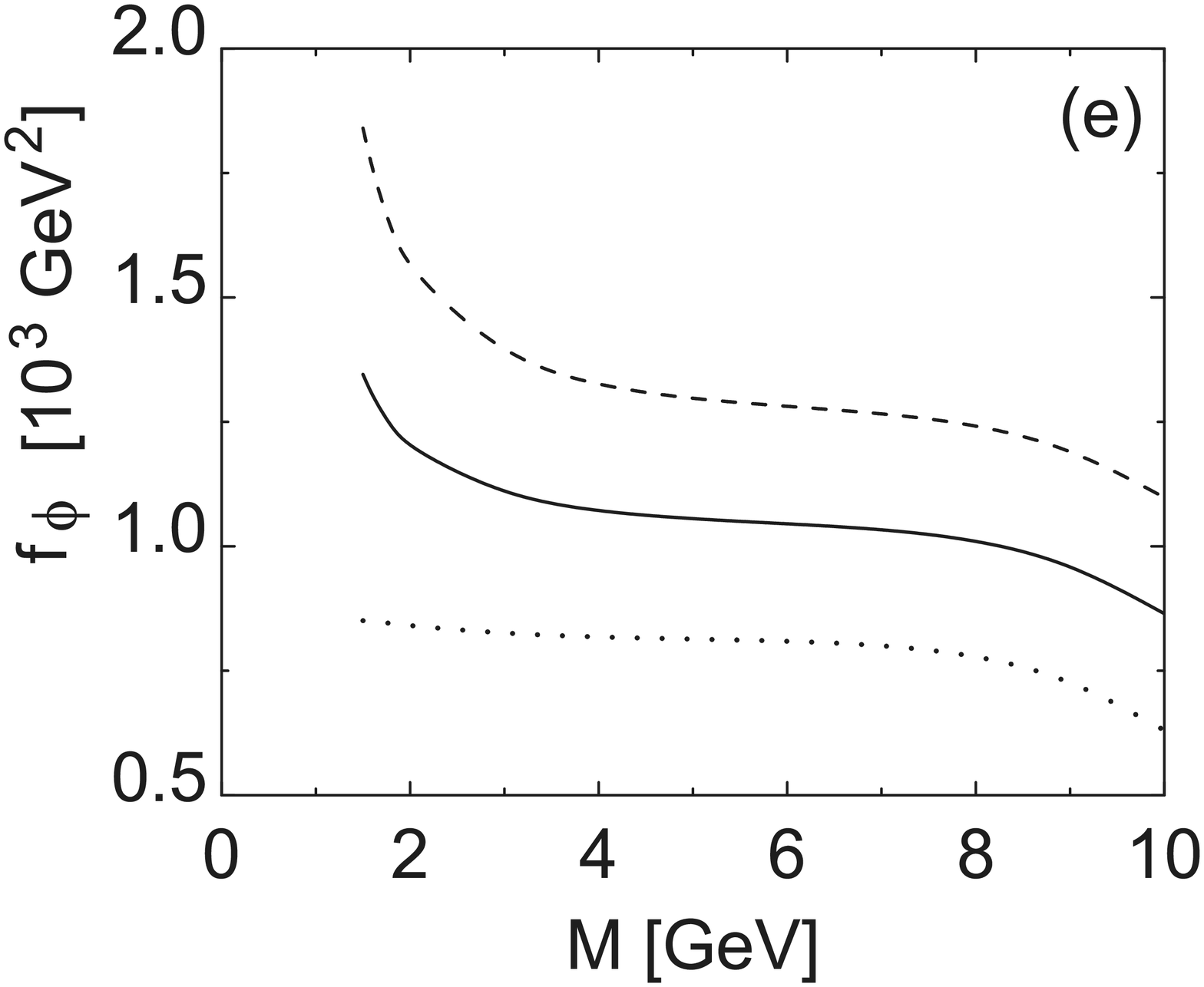}
\caption{Parameters $\zeta$ (a), $\beta$ (b), $\zeta'$ (c), 
$\beta'$ (d) and $f_{\phi}$ (e) 
as a function of the Borel mass.
Dotted curves are for vacuum, solid curves stand for $n=n_0$ and 
dashed curves depict $n=2\, n_0$. All curves are evaluated for $\kappa_N=
\kappa_0=3$ (here the mass shift of vector mesons has not been taken into 
account).}
\label{fig: figAA}
\end{figure}

In Fig.~\ref{fig: figAA} we have plotted these  
parameters as a function of the Borel mass for different densities. 
Like in the Borel analysis for the $\rho\; - \;\omega$ mass splitting we have 
to find an appropriate Borel window $M^2_{\rm min}, \; M^2_{\rm max}$.
To determine the minimal Borel window one could use again the $10\; \%$ rule 
getting $M_{\rm min} \approx 1$ GeV, while in \cite{Mixing15} a  
$25\;\%$ rule has been used getting $M_{\rm min} \approx 1.3$ GeV. 
But it turns out that in such a region 
around $M_{\rm min}$ the sum rule is 
unstable for a wide range of parameters \cite{Mixing15, Maltman}. 
Nevertheless, the curves in Fig.~\ref{fig: figAA}  
evidence that 
a stable region for all five unknowns exists in the interval 
$4 \le M \le 8$ GeV. 
This observation confirms a corresponding stability investigation in 
\cite{Maltman}. Therefore, in line with \cite{Maltman}, we will use a 
static Borel window $M_{\rm min} = 4 \le M \le M_{\rm max} = 8$ 
GeV over which we have to average (using eq.~(\ref{parameter_15})) to get Borel 
mass independent quantities. 
The result found in \cite{Mixing15,Maltman,Leinweber} that the threshold 
parameter $s_0$ in vacuum turns out to play a subdominant rule is also 
valid in case of finite density. Accordingly, we may use a fixed value, 
$s_V = 2.0$ GeV, for all densities.

\newpage

\subsection{Results}

\subsubsection{Vacuum}

\begin{figure}[!h]
\begin{center}
\includegraphics[scale=0.32]{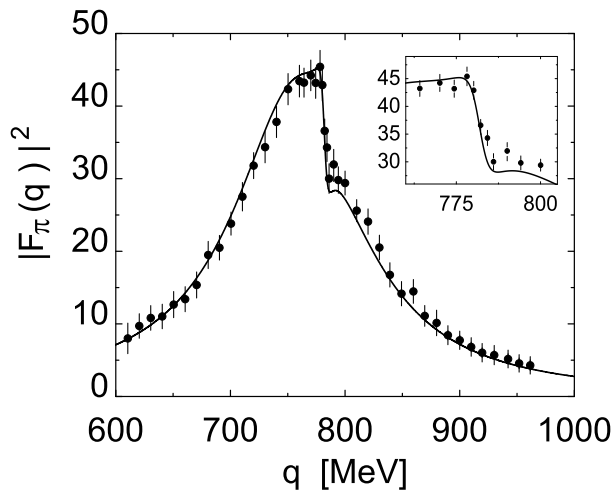}
\caption{Comparision of the formfactor evaluated within the  
QCD sum rule method (solid curve) and the results of the CMD-2 experiment 
(symbols) \cite{experiment_formfactor}.}
\label{fig: fig7}
\end{center}
\end{figure}

First of all, let us briefly discuss the pion formfactor in vacuum, 
given by (\ref{mixing_30}) with $n=0$. 
Our sum rule analysis for vacuum results in  
$\zeta=1.055 \,\times\,10^{-3}$ in good agreement with \cite{Mixing15,Mixing5}. 
Using (\ref{epsilon}) gives for the mixing parameter  
$\epsilon = - 0.21$. 
The hadronic contribution of the nondiagonal on-shell selfenergy 
which enters the pion formfactor, is given, via eq.~(\ref{mixing_sumrule_40}), 
by $\delta_{\rho\,\omega}^{\rm H} (\overline{m}^2) = - \overline{m}^2 
g_{\rho\,\gamma}\,g_{\omega\,\gamma}\, \zeta /12 = 
- 4289$ MeV$^2$ which amounts, by taking into account the electromagnetic 
nondiagonal on-shell selfenergy $\delta_{\rho\,\omega}^{\rm EM} 
(\overline{m}^2) = 610$ MeV$^2$ \cite{self_feynman,Mixing15}, in total to 
$\delta_{\rho\,\omega} (\overline{m}^2) = - 3679$ MeV$^2$, in fair agreement 
with experiment \cite{Pionformfactor_5}. 
Using this value we get the pion formfactor in vacuum as shown in 
Fig.~\ref{fig: fig7}. 
It reproduces very well recently obtained experimental 
data \cite{experiment_formfactor}. In \cite{Maltman} it was argued 
that the sum rules might give not a good agreement with data when taking 
the parameter set of \cite{Mixing15}. Our analysis shows, however, that 
the sum rules are in agreement with experimental data when using  
appropriate parameters. 

\subsubsection{Isospin symmetric nuclear matter}

\begin{figure}[!h]
\includegraphics[scale=0.2]{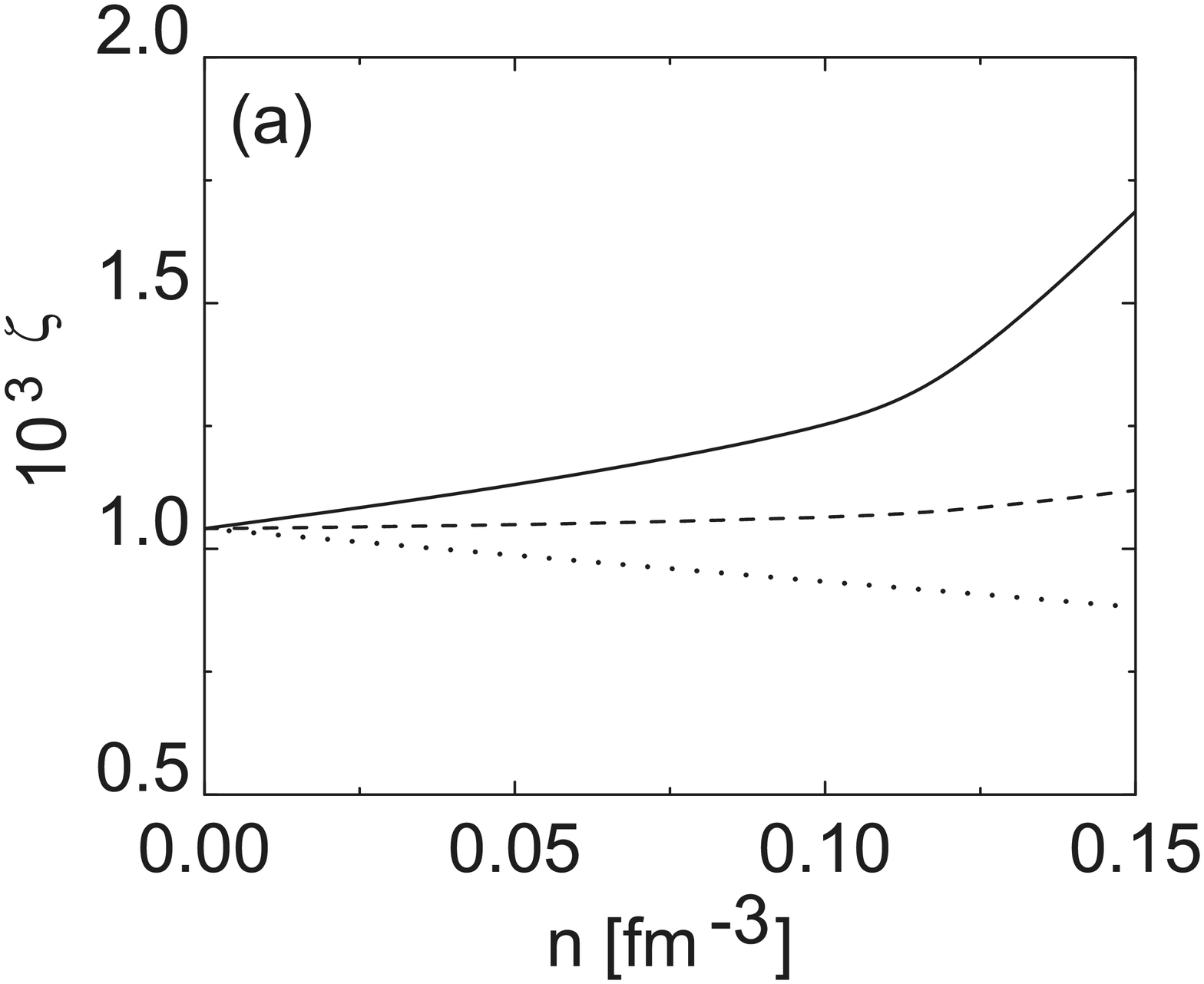}
\hspace{-0.5cm}
\includegraphics[scale=0.2]{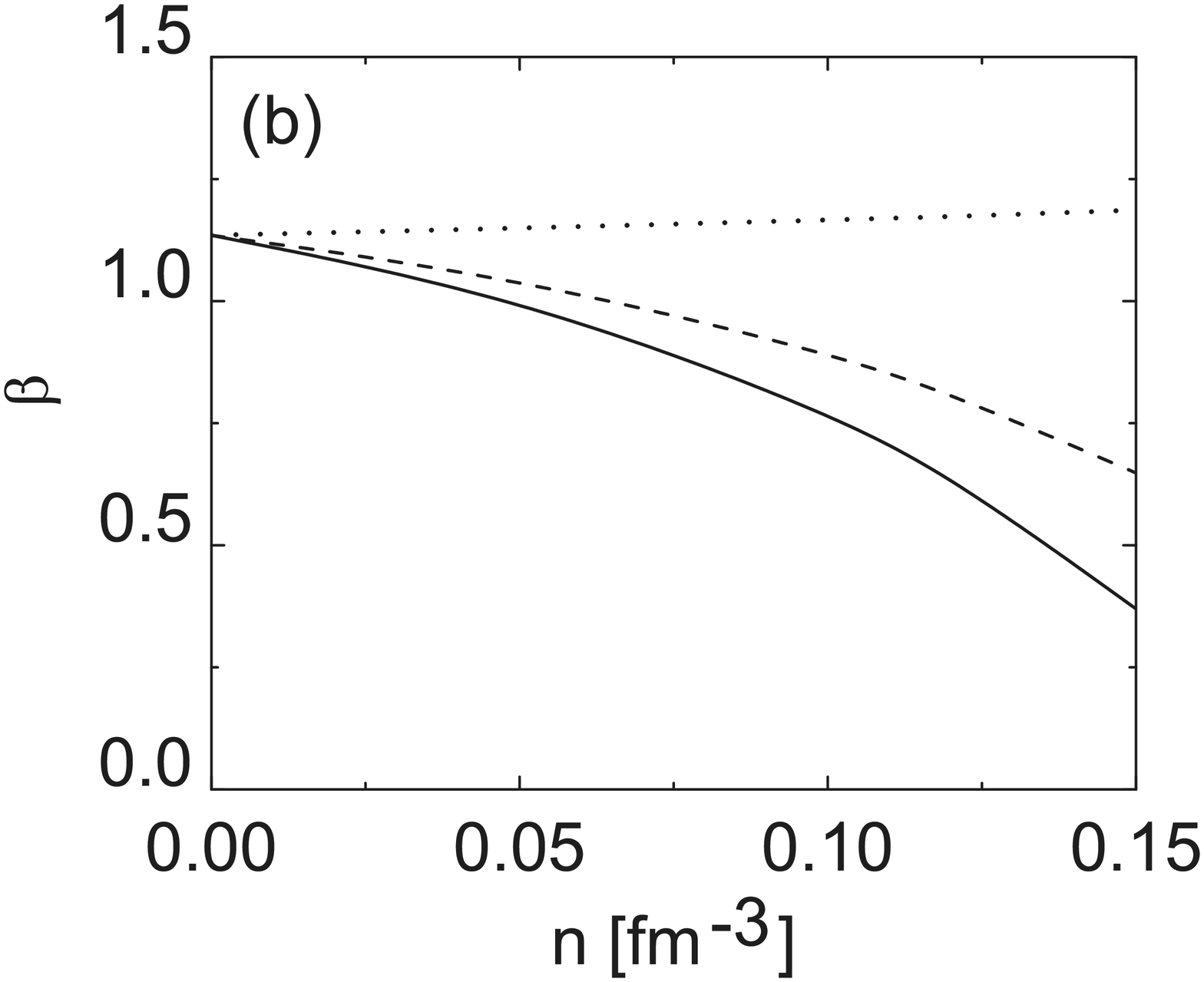}
\hspace{-0.5cm}
\includegraphics[scale=0.2]{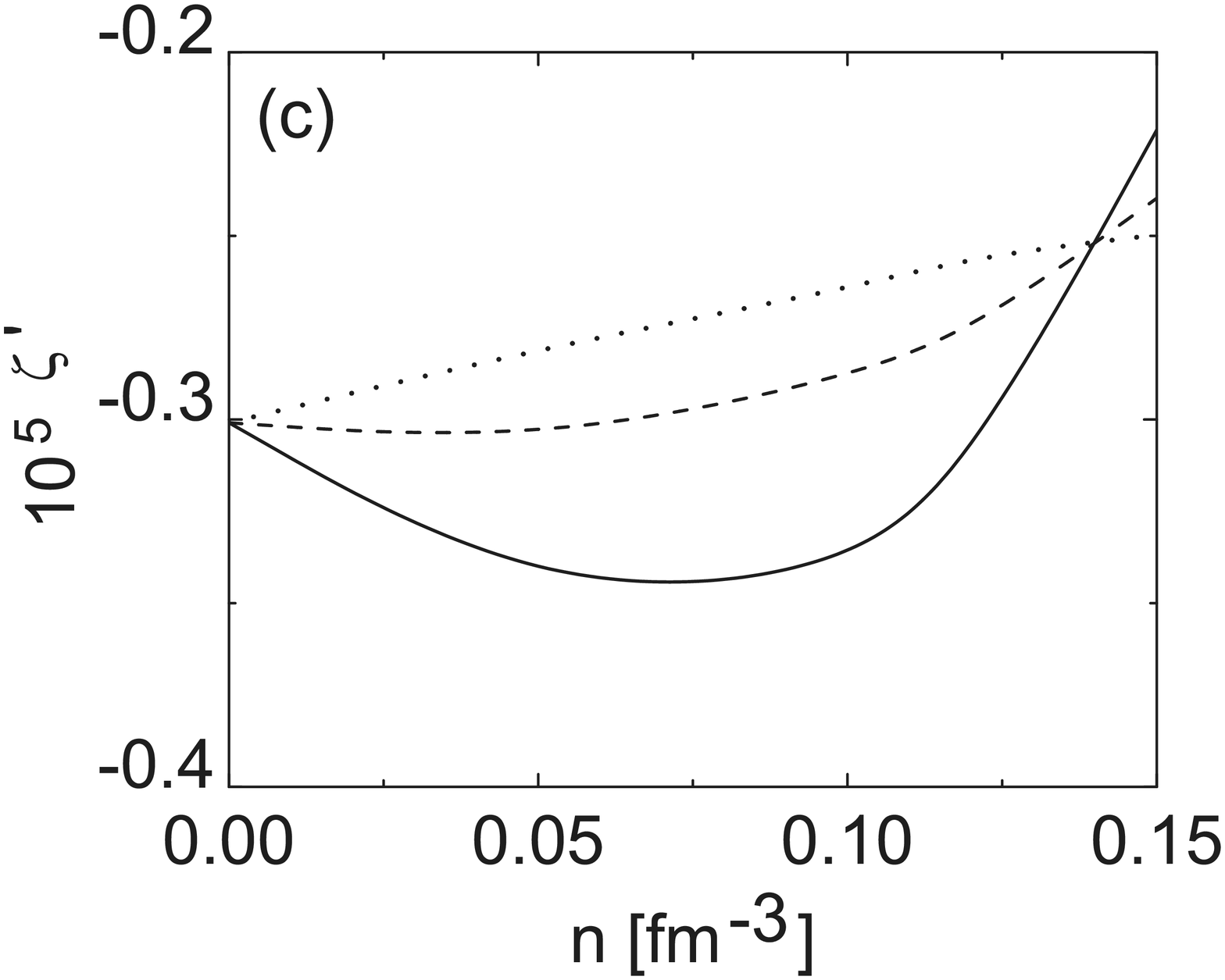}
\hspace{-0.5cm}
\includegraphics[scale=0.2]{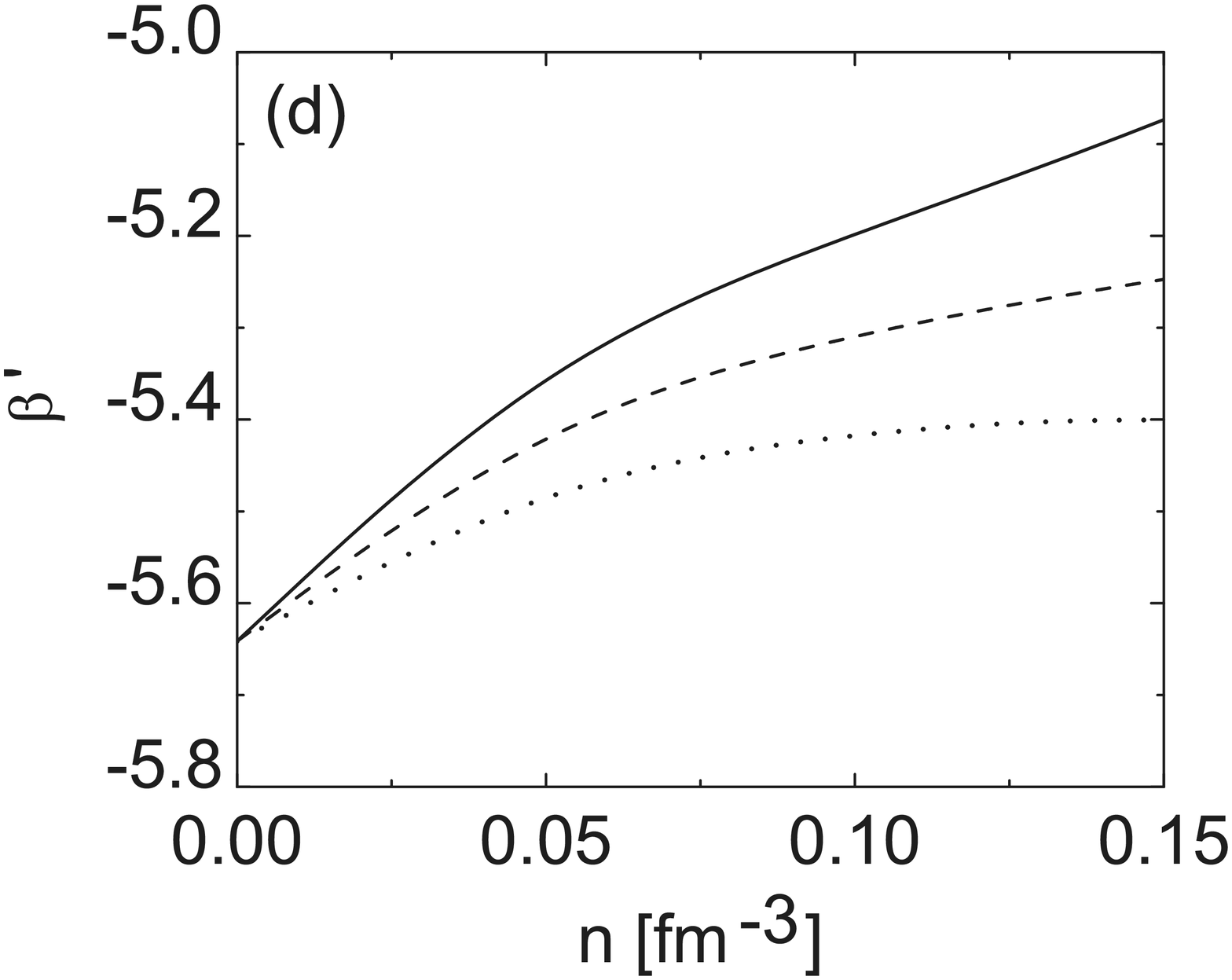}
\hspace{-0.5cm}
\includegraphics[scale=0.2]{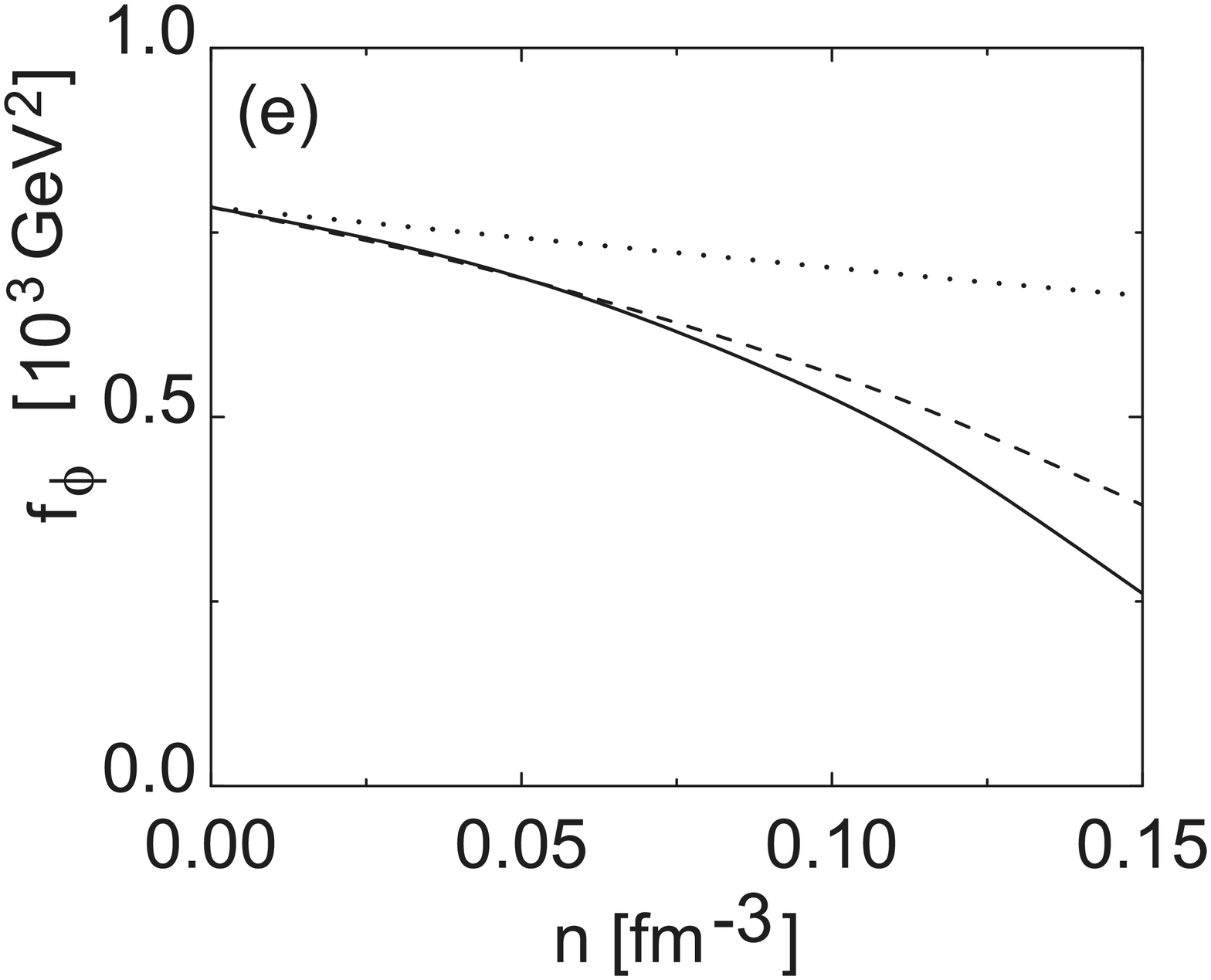}
\caption{Parameters $\zeta$ (a), $\beta$ (b), $\zeta'$ (c), $\beta'$ (d) 
and $f_{\phi}$ (e)
at finite density. Dotted curves are for $\kappa_N=2$, dashed curves are for 
$\kappa_N=3$ and solid curves are for $\kappa_N=4$. 
The density dependence of the mass parameters of $\rho$, $\omega$ and $\phi$
mesons  
(without the twist-4 condensates) has been taken into account consistently.}
\label{fig: figAAA}
\end{figure}

After reproducing the pion formfactor in vacuum we now turn 
to the density dependence of the mixing effect.  
Due to the small effect of mixing compared to splitting 
and the large impact of the four-quark condensate and Landau damping terms 
on mass parameter  
splitting, it becomes obvious that mixing does not strongly influence the mass
parameter splitting effect. 
But on the other side, the mass parameter splitting effect could
strongly influence the mixing effect. 
To study the effect of the $\rho - \omega$ mass parameter splitting on the 
$\rho - \omega$ mixing we have to implement in the five equations for 
the five unknowns $\zeta,\beta,\zeta',\beta',f_{\phi}$ the density 
dependent mass parameters, i.e.: $m_{\rho} (n)$, $m_{\omega} (n)$
and $m_{\phi} (n)$, respectively. For $m_{\rho} (n)$, $m_{\omega} (n)$ 
we use the values obtained in the previous section, while for 
the density dependence of the $\phi$ meson we will take  
the relation $m_{\phi} (n) = (1 - \alpha \;n/n_0) m_{\phi} (0)$ with 
$\alpha = 0.03$, which turns out to be almost independent of $\kappa_N$ 
\cite{zschocke1}.
The results for the five parameters are shown in Fig.~\ref{fig: figAAA}.

\begin{figure}[!h]
\includegraphics[scale=0.3]{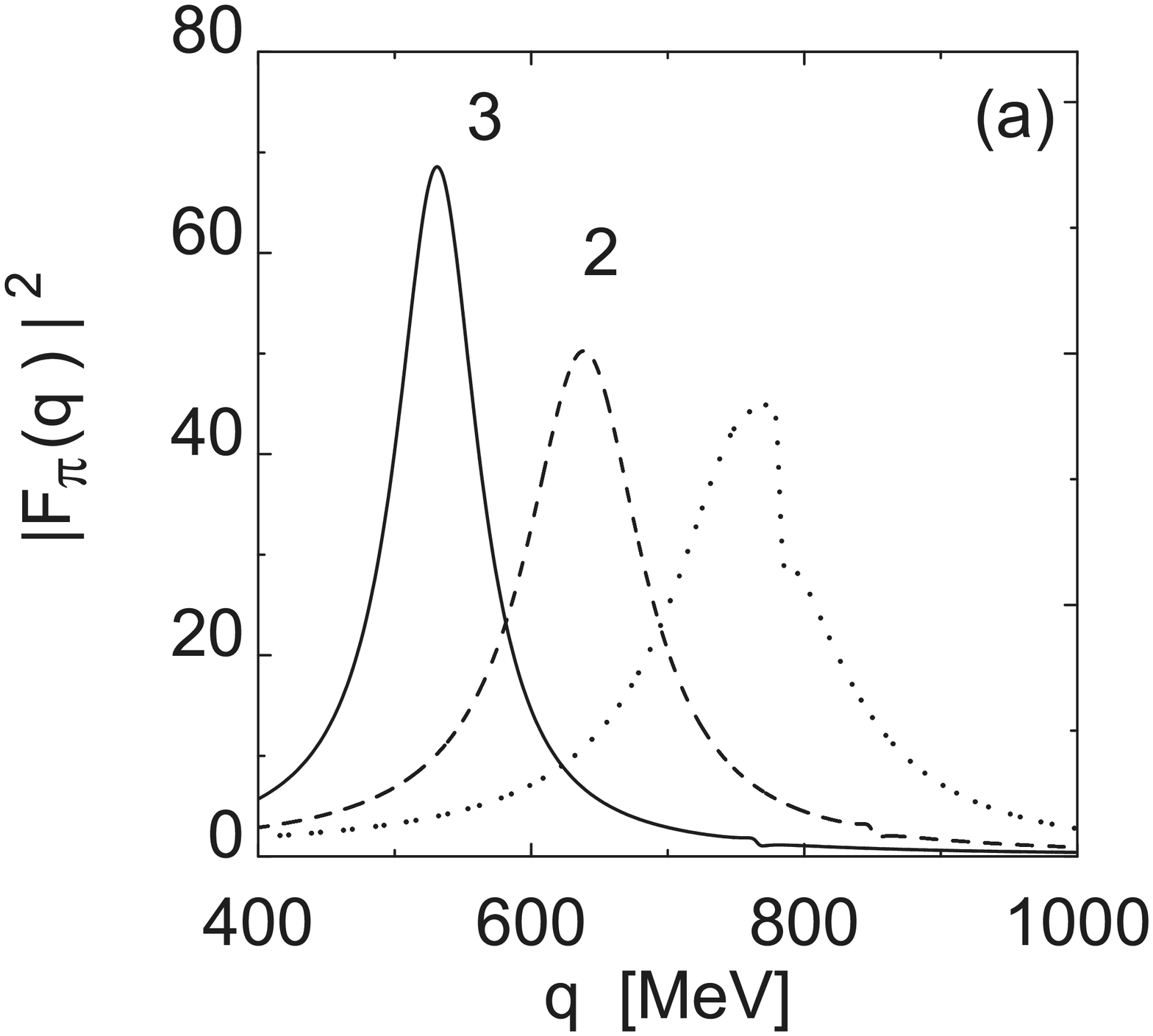}
\hspace{0.0cm}
\includegraphics[scale=0.3]{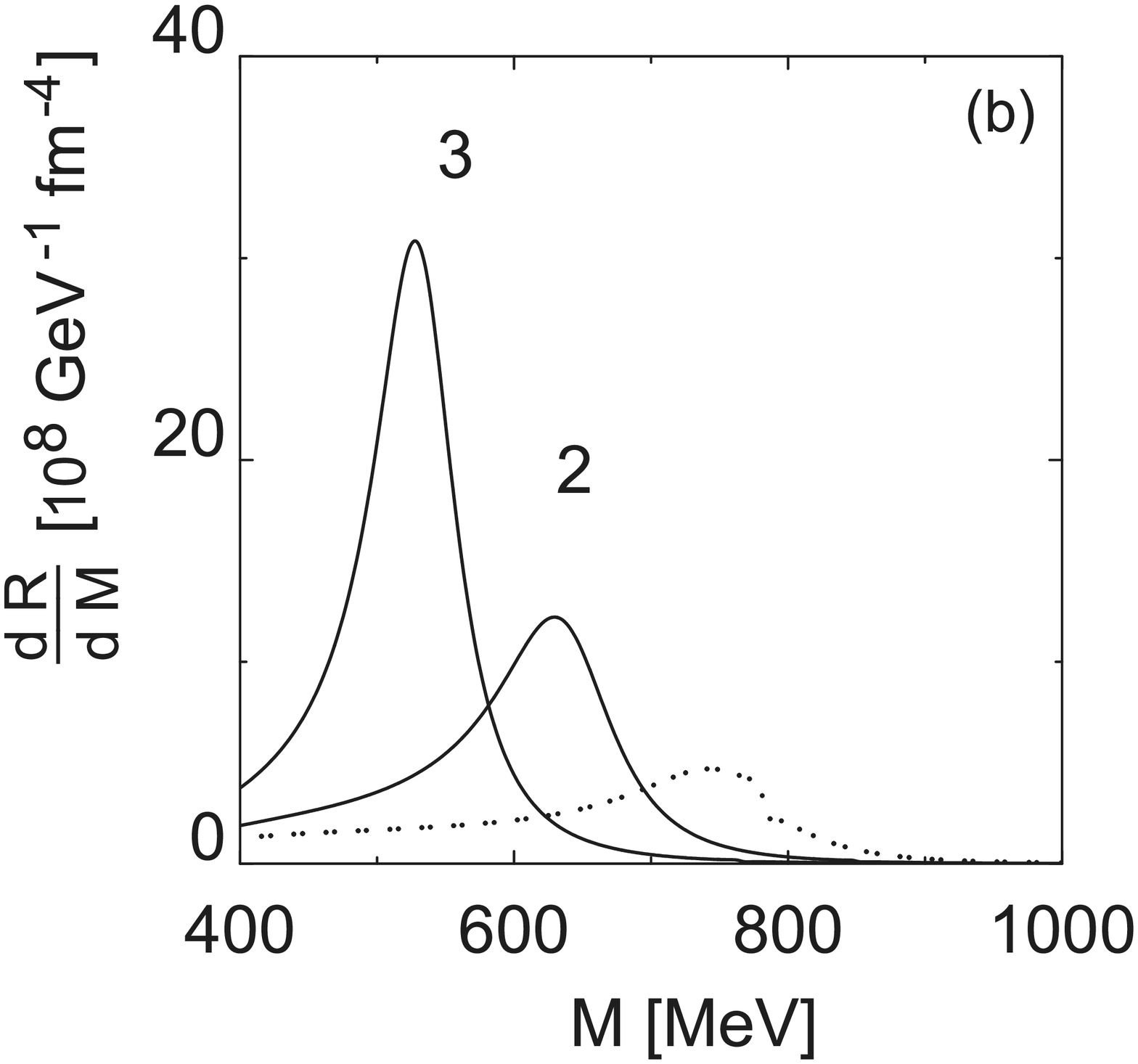}
\caption{Left pannel (a): Pion formfactor at saturation density $n=n_0$.
Mass shifts of vector mesons (without twist-4 condensates) are taken into 
account. The dotted curve is for vacuum, while
the solid curves are for saturation density $n=n_0$.
The labels $1,2,3$ denote $\kappa_N=1,2,3$, respectively.
Right pannel (b): Di-electron production rate from pion-pion annihilation 
at finite density and $T=100$ MeV.
The dotted curve is for a hot pion gas and baryonic 
vacuum $n=0$, while the solid curves are for saturation density $n=n_0$.}
\label{fig: fig_Mix_Splitt}
\end{figure}

From the density behavior of the parameter $\zeta$ 
(see Fig.~\ref{fig: figAAA} (a)) one might  
conclude that the mixing effect remains in matter. 
But this is actually not the case. In view of eq.~(\ref{epsilon}) we recognize 
that the mixing angle $\epsilon$ is strongly suppressed by the factor
$1/\Delta m^2$. Additionally, the mass shift of the $\rho$ meson
modifies significantly the pion formfactor.
Using eqs.~(\ref{mixing_30}) and (\ref{mixing_sumrule_65})
for the pion formfactor and di-electron production rate, respectively,
we get the results shown in Fig.~\ref{fig: fig_Mix_Splitt}.

These figures show that the mixing effect in the pion formfactor as well as
in the di-electron production rate is washed out due to the
mass shifts of the vector mesons.
But one has to keep in mind that global changes of vector mesons in matter
like mass shift and width broadening turn out to be correlated in nuclear
matter \cite{Leupold2,Leupold3,zschocke3}. Taking into account such  
broadening effects needs further investigations.

\begin{figure}[!h]
\includegraphics[scale=0.2]{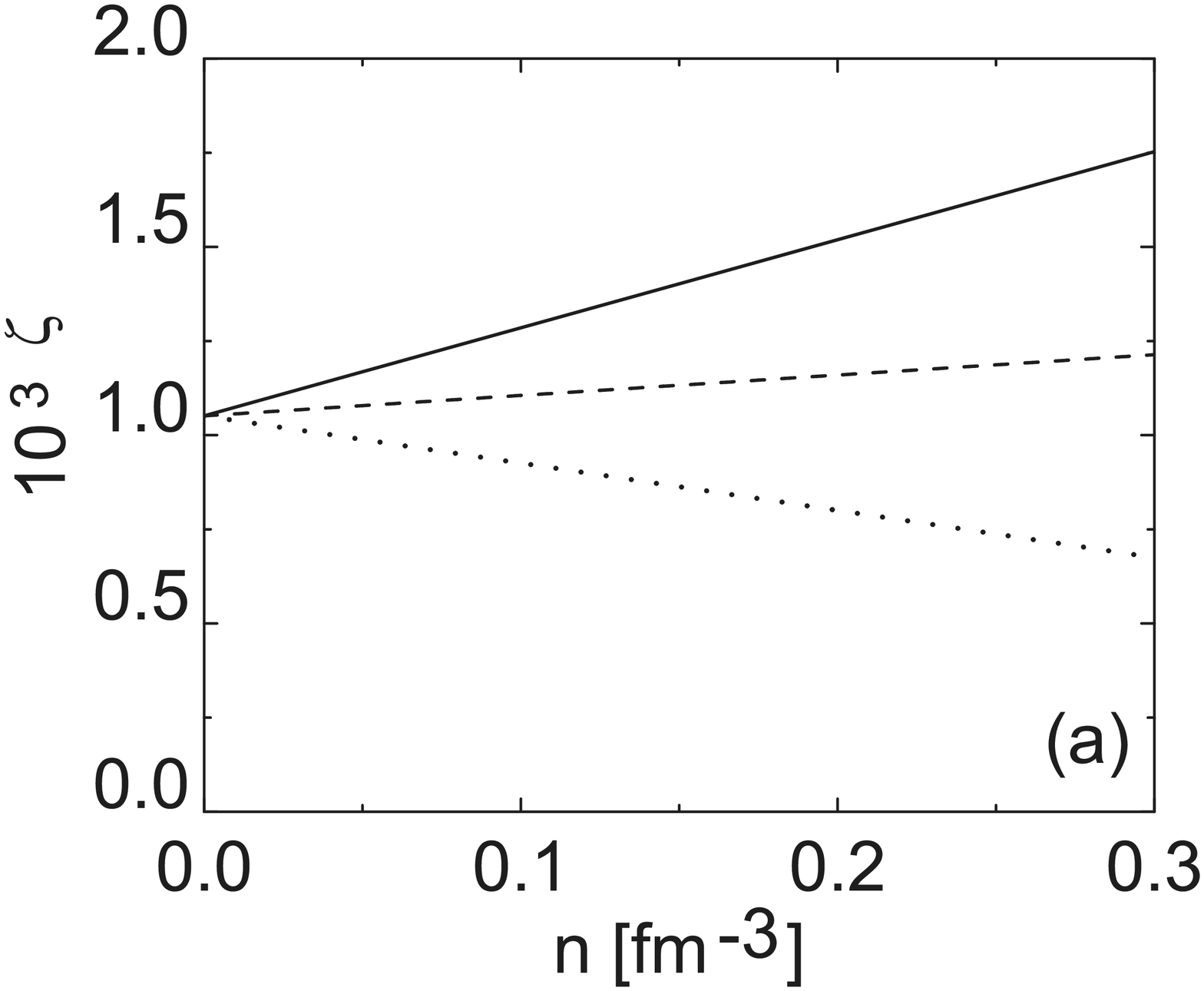}
\hspace{-0.5cm}
\includegraphics[scale=0.2]{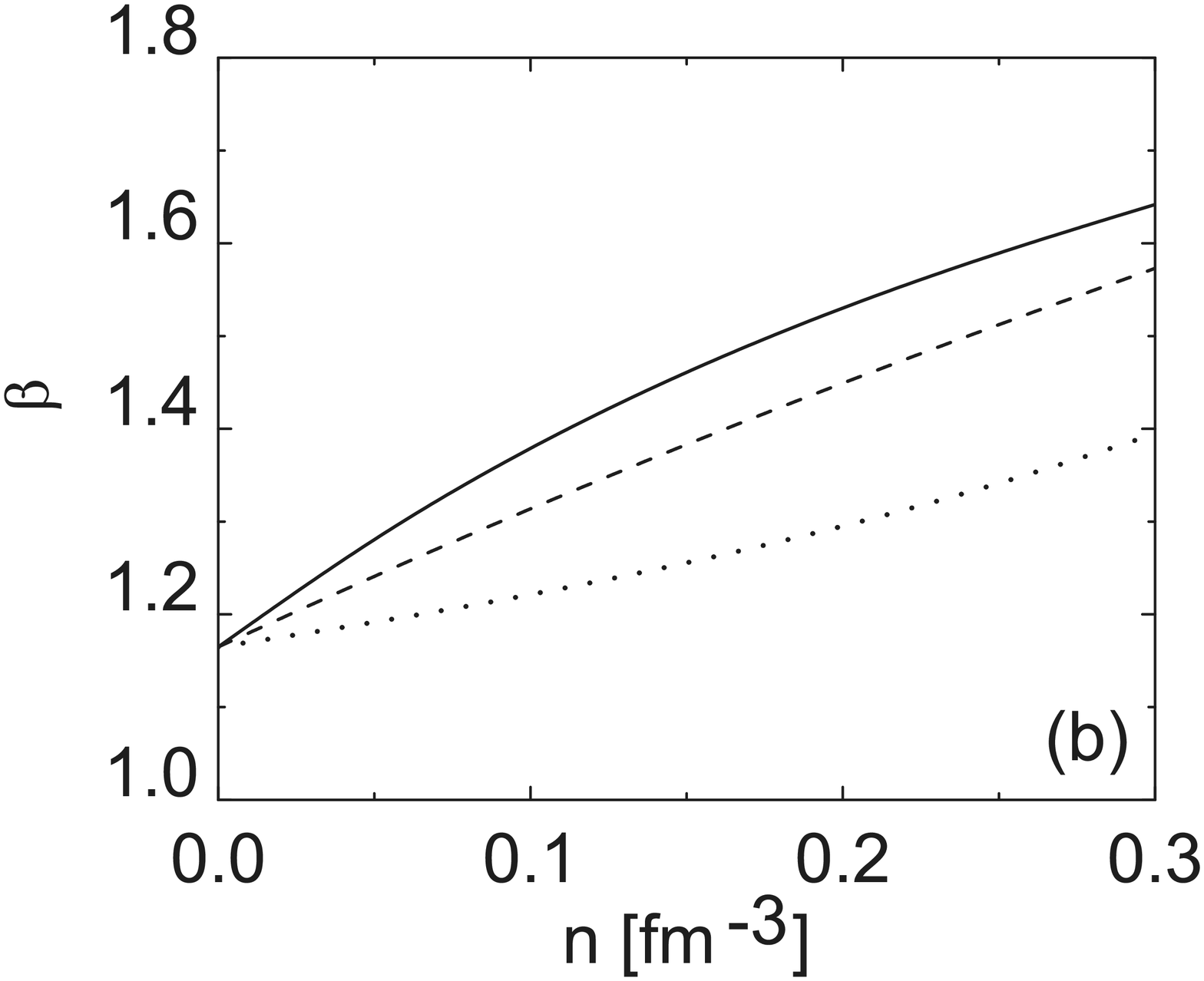}
\hspace{-0.5cm}
\includegraphics[scale=0.2]{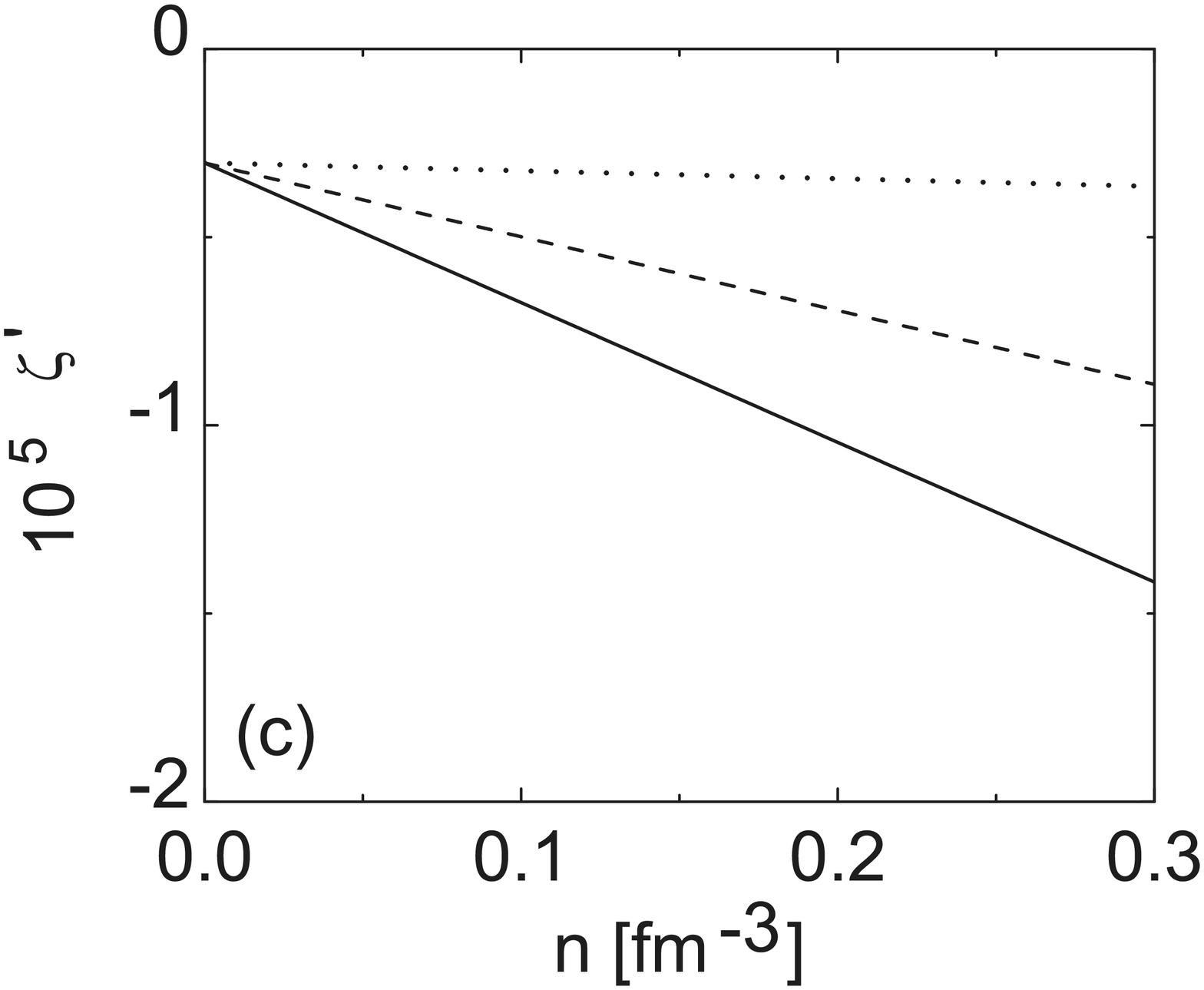}
\hspace{-0.5cm}
\includegraphics[scale=0.2]{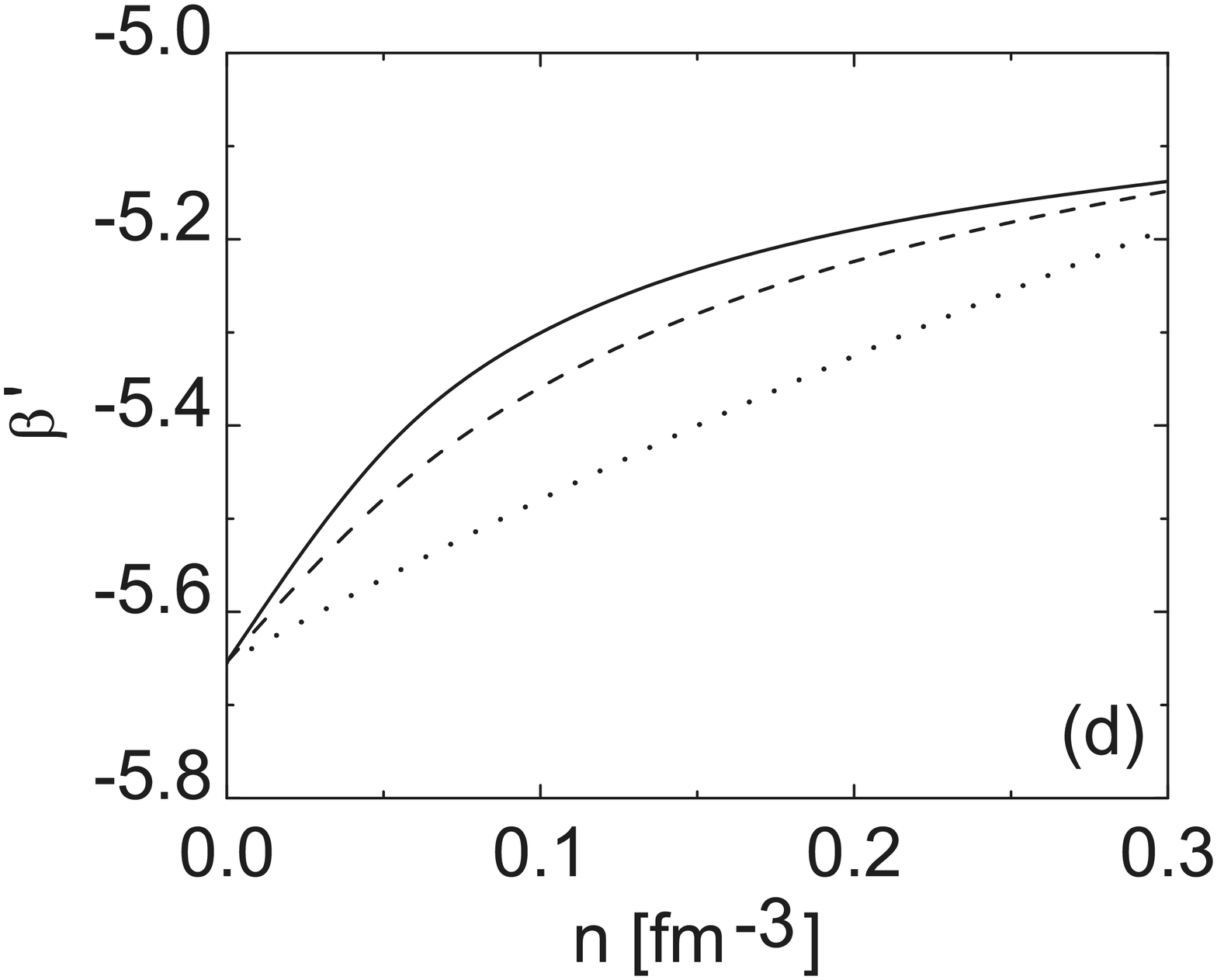}
\hspace{-0.5cm}
\includegraphics[scale=0.2]{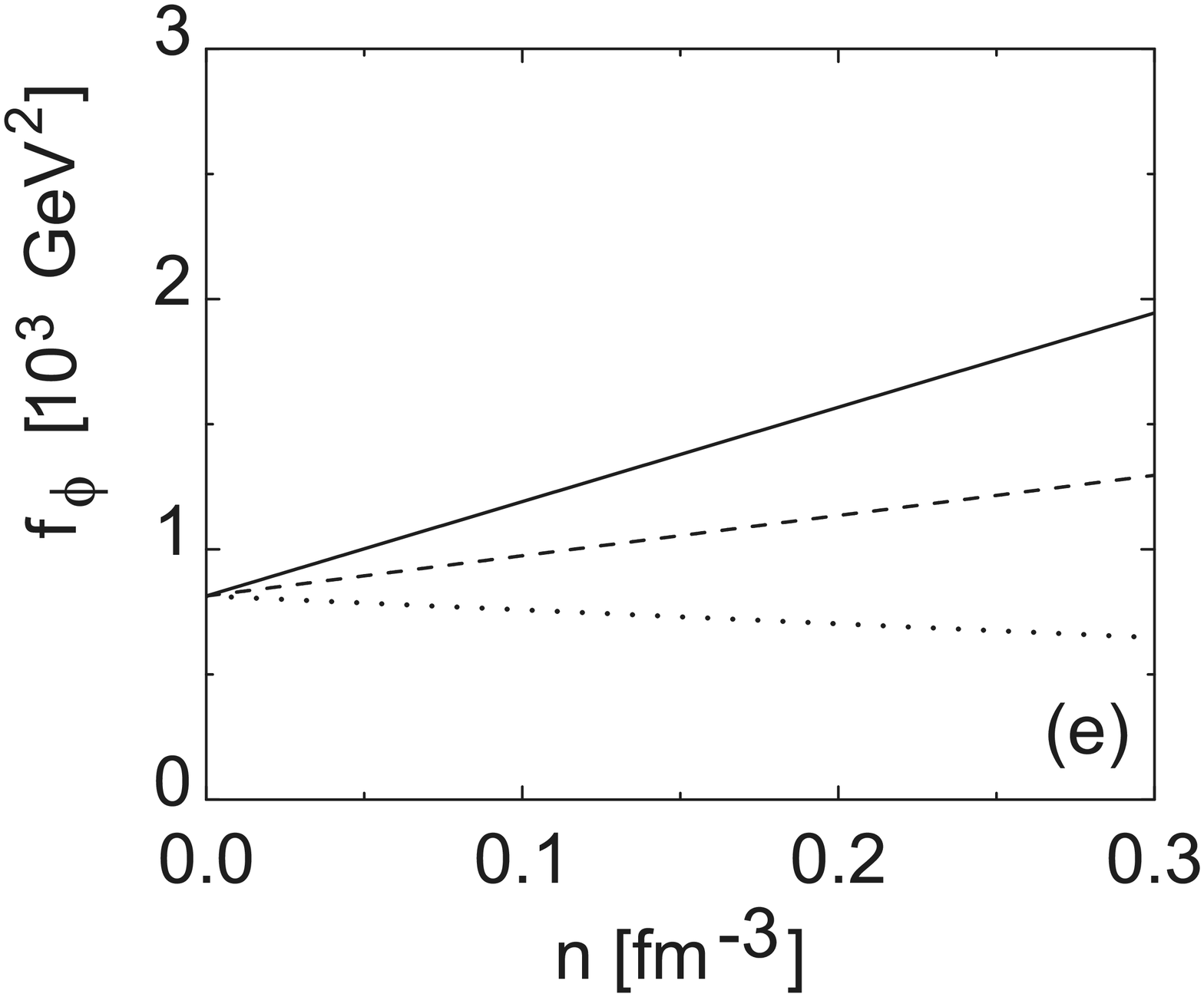}
\caption{Parameter $\zeta$ (a), $\beta$ (b), $\zeta'$ (c), $\beta'$ (d) 
and $f_{\phi}$ (e) 
at finite density. The density dependence of the mass parameters has not been 
taken into account. Same notation as in Fig.~\ref{fig: figAAA}.}
\label{fig: figA}
\end{figure}

\begin{figure}[!h]
\includegraphics[scale=0.3]{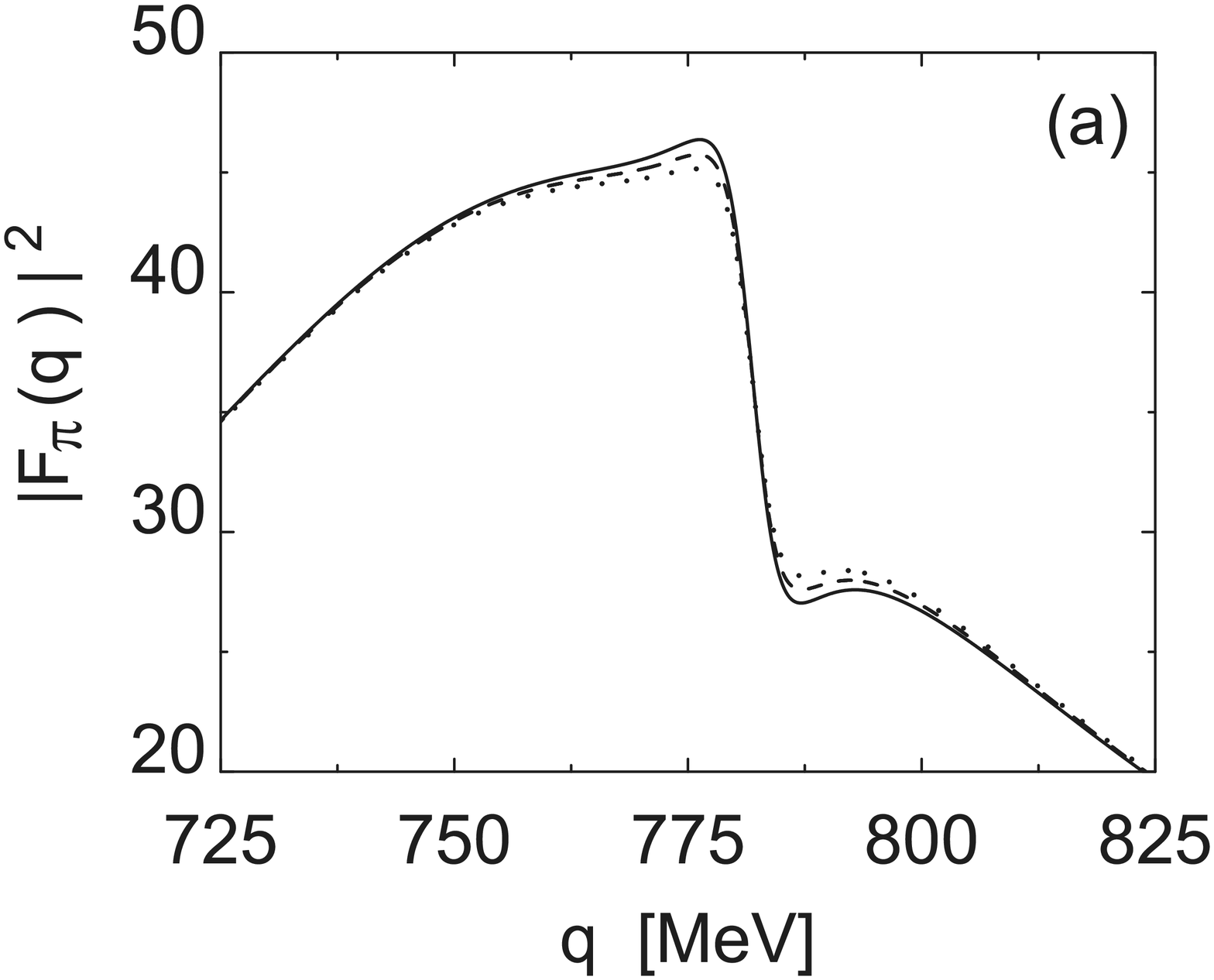}
\hspace{0.0cm}
\includegraphics[scale=0.3]{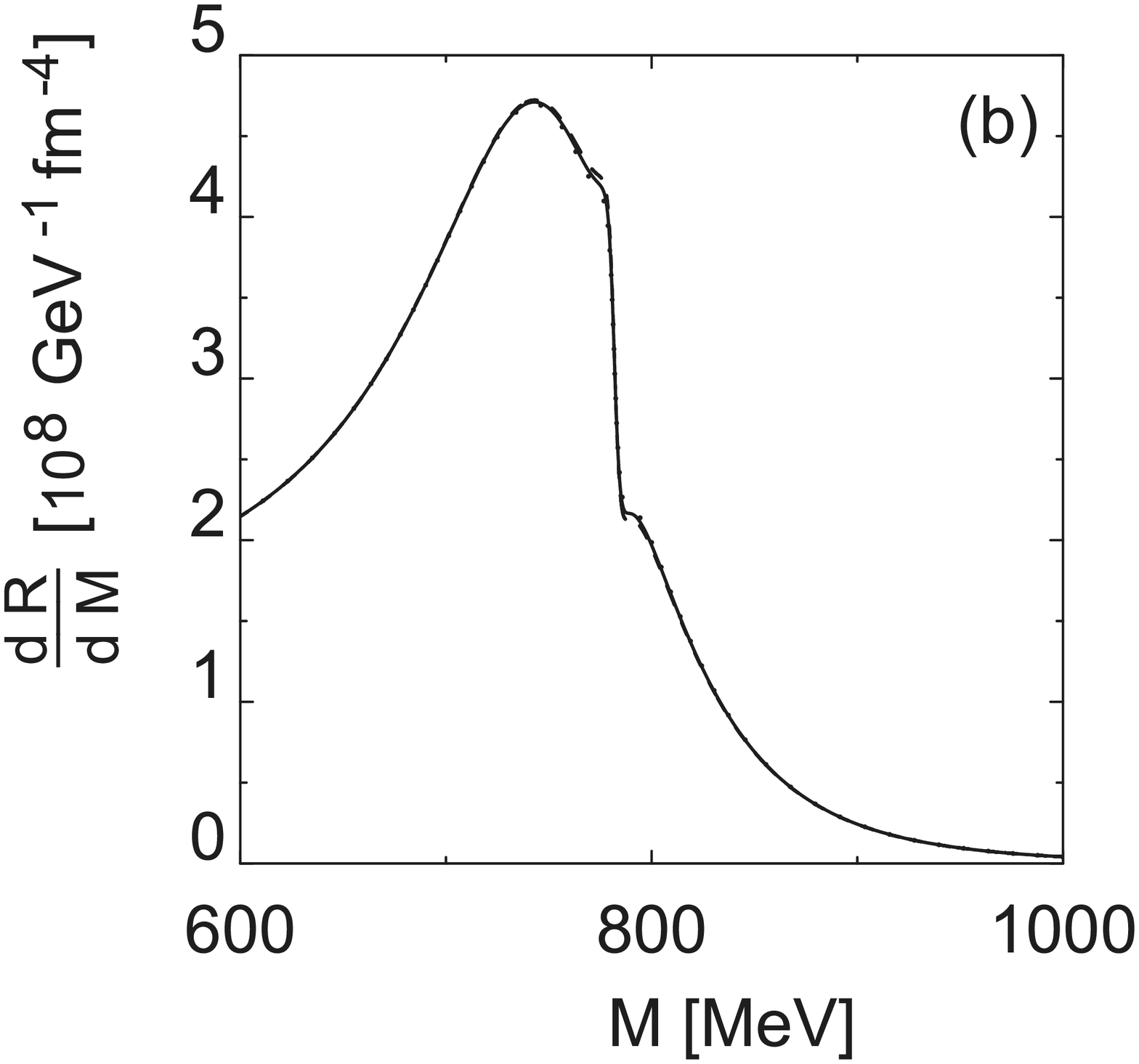}
\caption{Left pannel (a): Formfactor at finite density. Dotted line denotes 
vacuum, 
dashed line represents $n=n_0$ and solid line means $n=2 n_0$.
Right pannel (b): Di-lepton production rate for pion-pion annihilation
at finite density for $T=100$ MeV. The plotted curves are for  
$\kappa_N=3$. No mass shifts} 
\label{fig: fig8}
\end{figure}

We also show results without the mass shifts  
of $\rho$, $\omega$ and $\phi$ mesons.
The corresponding density dependence of the five parameters 
$\zeta,\beta,\zeta^{'},\beta^{'},f_{\phi}$ is shown in Fig.~\ref{fig: figA}. 
One observes noticeable changes for the parameters. 
The dashed curve (i.e. $\kappa_N = \kappa_0$)
in Fig.~\ref{fig: figA} (a)
recovers the density-independence of $\zeta$ for  
isospin symmetric nuclear matter as anticipated in \cite{Mixing5}. 
Otherwise, depending on the parameter $\kappa_N$ which governs
the density dependence of the four-quark condensate, $\zeta$ may slightly 
increase (large $\kappa_N$) or decrease (smaller $\kappa_N$) with increasing 
density. The resulting 
pion formfactor and the di-electron production rate are plotted in 
Fig.~\ref{fig: fig8}. 
At finite density one obtains a very small modification of the formfactor 
compared to the vacuum, while the modification of the rate is nearly invisible.  

In the Figs.~\ref{fig: fig_Mix_Splitt} and \ref{fig: fig8} 
the meson peaks are assumed to be distributed with a schematic width 
${\rm Im} \Sigma_{\rho} (E)= - \frac{g_{\rho \pi \pi}^2}
{(48 \pi) E }(E^2 - 4 m_{\pi}^2)^{3/2}\; \Theta (E- 2 m_{\pi})$
and ${\rm Im} \Sigma_{\omega} (E) =
- m_{\omega} \Gamma_{\omega} \,\Theta (E- 3 m_{\pi})$, respectively.  
In Fig.~\ref{fig: fig_Mix_Splitt} the density dependence of  
$m_{\rho,\omega}$ are taken into account, while in Fig.~\ref{fig: fig8}   
no shifts of $m_{\rho,\omega}$ are assumed.  
Obviously, the di-electron rates shown in Figs.~\ref{fig: fig_Mix_Splitt}
and \ref{fig: fig8} differ significantly.
 
There is the possibility, advocated in 
\cite{Mosel}, that in-medium the original (vacuum) $\rho$ peak is not shifted, 
but additional strengths develops below the $\rho$ peak. A similar 
possibility has been reported in \cite{TVN1} for the $\omega$ 
meson. In such cases the $\rho - \omega$ mixing remains, similar to 
Fig.~\ref{fig: fig8}, but the weighted $\rho$ strength is shifted 
down, as required by the sum rule considered in section 2. A proper handling 
of this situation deserves further investigations with explicit knowledge 
of the $\rho$ and $\omega$ in medium spectral functions. 
Experimentally, precision measurements with HADES \cite{HADES} can deliver 
informations on the in-medium behavior of the $\rho - \omega$ mixing.

\subsubsection{Isospin asymmetric nuclear matter}

So far we have considered isospin symmetric nuclear matter.  
While it is not necessary to study isospin asymmetric matter 
for the mass splitting effect, finite values of $\alpha_{n p}$ have 
some relevance for the mixing effect \cite{Mixing5}. Therefore, in  
this subsection we concentrate on isospin asymmetric nuclear matter.  
The needed proton and neutron condensates are given in the Appendix E.  
Accordingly, the coefficients $d_i^{AS}$ in lines (\ref{mixing_sumrule_29}) 
and  (\ref{mixing_sumrule_30}) contain the following  
terms proportional to $\alpha_{n p}$:  
\bea
d_2^{\rm AS} &=& - \frac{1}{2}\; \left(1 + \frac{\alpha_s}{\pi}\; C_F\; 
\frac{1}{4} \right)\; \frac{n}{2 M_N} m_q \;\alpha_{n p} 
\;\langle p | \overline{u} u - \overline{d} d | p \rangle\nonumber\\
\nonumber\\
&& - \frac{1}{72}\; \frac{\alpha_{\rm em}}{\pi}\; \frac{1}{M_N}\; 
\frac{n}{4} \; \alpha_{n p}\; 5 \; m_q \; 
\langle p | \overline{u} u - \overline{d} d | p \rangle 
\nonumber\\
\nonumber\\
&& - \left(\frac{1}{4} - \frac{5}{48}\;\frac{\alpha_s}{\pi} \; C_F \right) 
\; \left(A_2^{u, p} - A_2^{d, p} \right) \; M_N \; \alpha_{n p}\; n 
\nonumber\\
\nonumber\\
&& + \frac{\alpha_{\rm em}}{\pi} \frac{25}{864}\; M_N \;n\;\alpha_{n p} 
\left(A_2^{u, p}
- A_2^{d, p} \right)\;,
\label{asymmetric5}\\
\nonumber\\
d_3^{\rm AS} &=& \frac{56}{81} \; \pi\;\alpha_s \;\alpha_{n p} \;n\; 
\frac{1}{M_N}\; \langle \overline{q} q \rangle_0\; 
\langle p | \overline{u} u - \overline{d} d | p \rangle 
\nonumber\\
\nonumber\\  
&& + \frac{7}{81}\; \pi\;\alpha_{n p} \; n\; \frac{1}{M_N}\; \kappa_N\; 
\langle \overline{q} q \rangle_0\;
\langle p | \overline{u} u - \overline{d} d | p \rangle
\nonumber\\
\nonumber\\
&& + \frac{\alpha_{\rm em}}{\pi} \frac{335}{3456}
M_N^3 \;n \;\alpha_{n p}\;\left(A_4^{u, p} - A_4^{d, p} \right) \;,
\label{asymmetric10}
\eea
where terms of order ${\cal O}\left( (m_d - m_u) \alpha_{n p} \right)$,  
${\cal O}\left(\gamma\,\alpha_{\rm em} \right)$ 
and ${\cal O}\left(\gamma\,\alpha_{n p} \right)$ are neglected. 

\begin{figure}[h]
\includegraphics[scale=0.3]{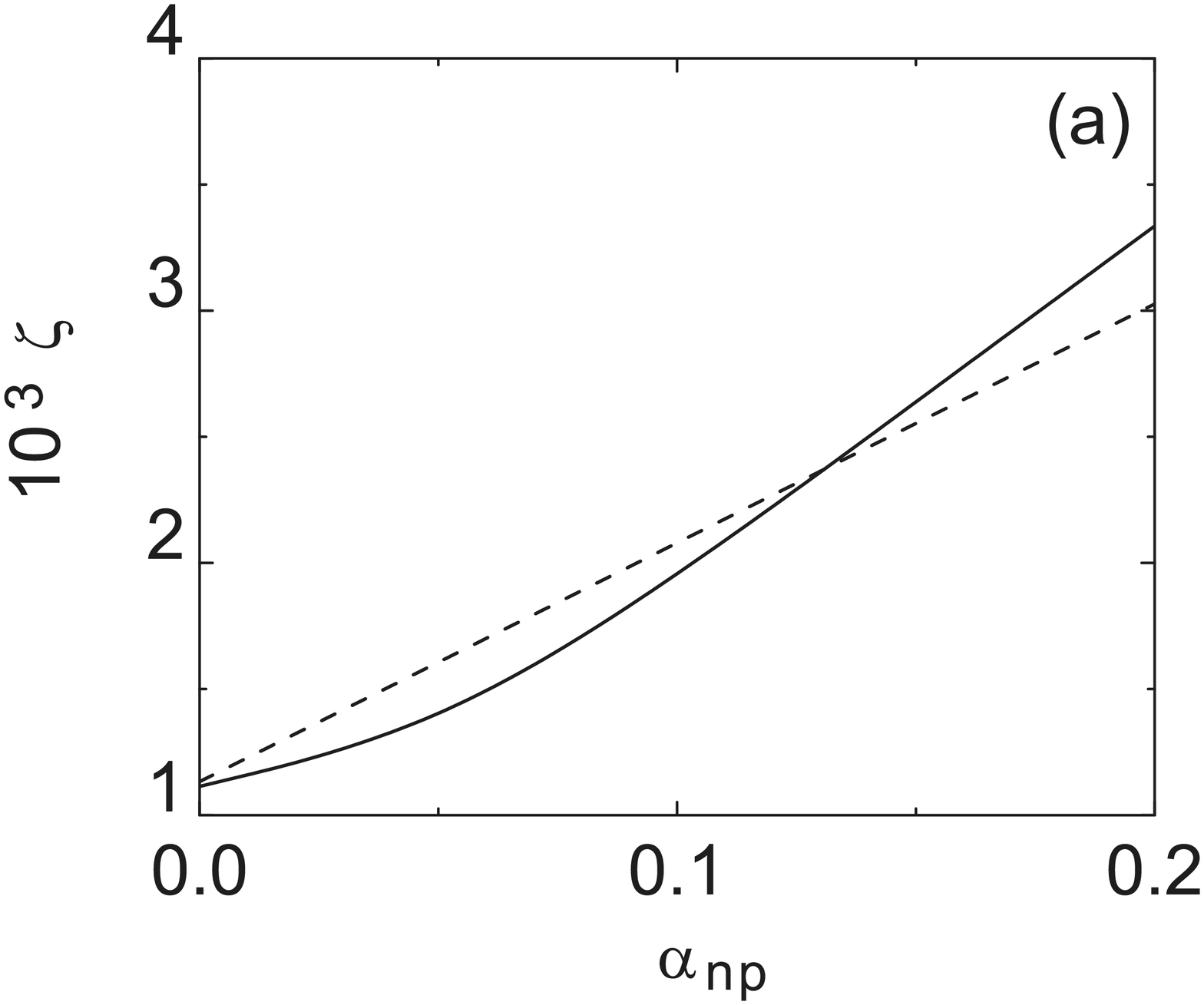}
\hspace{-0.5cm}
\includegraphics[scale=0.3]{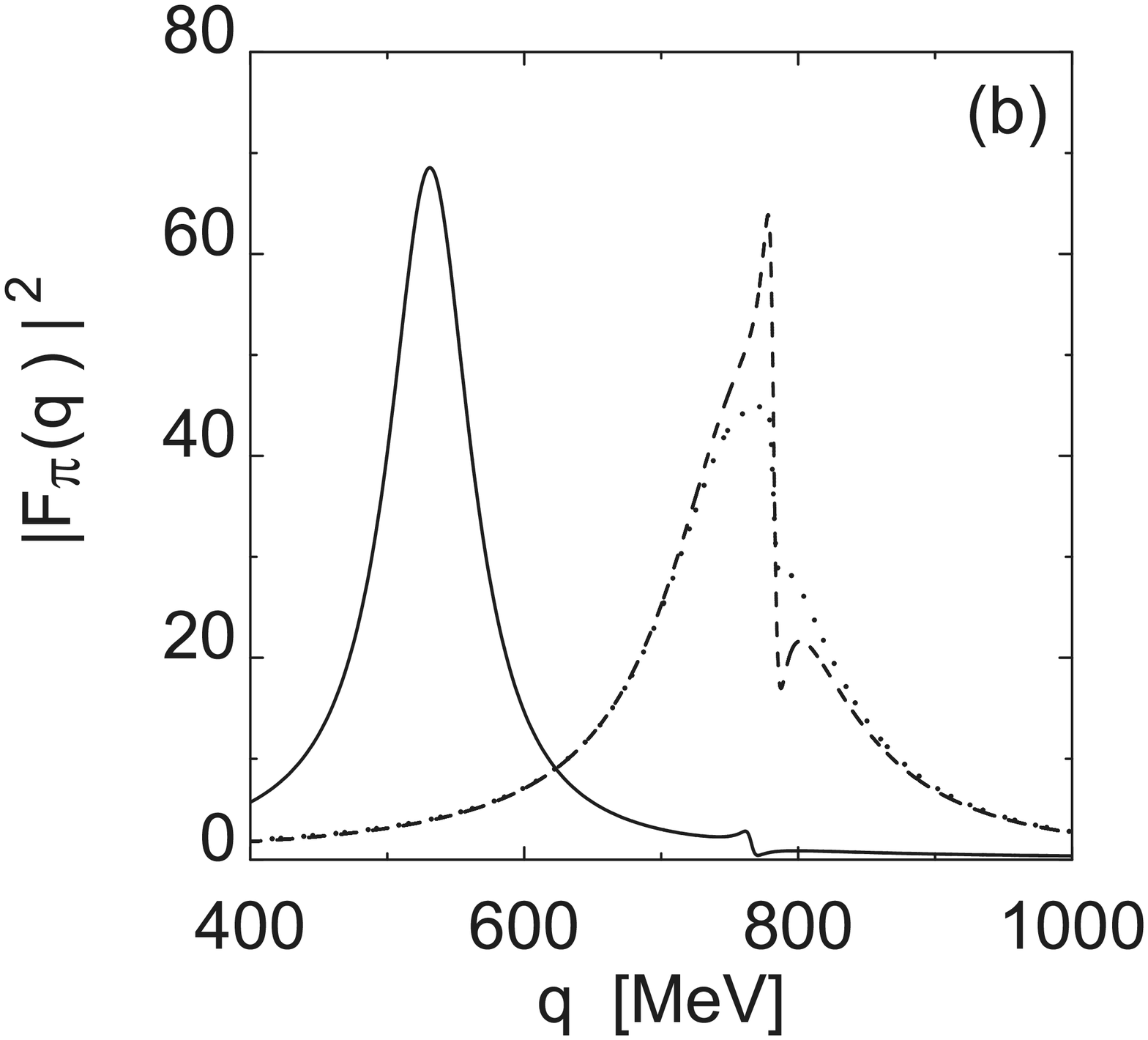}
\caption{Left pannel (a): Parameter $\zeta$ as a function of 
$\alpha_{n p}$ at 
saturation density $n_0$ (solid line: individual mass shifts of vector 
mesons have been taken into account; dashes line: without mass shifts 
of vector mesons).  
Right pannel (b): Pion formfactor for $\kappa_N=3$ and 
$\alpha_{n p} = 0.2$ 
(dotted line: vacuum; solid line: $n=n_0$ with mass shifts of vector mesons; 
dashed line: $n=n_0$ without mass shifts of vector mesons).}
\label{fig: Asymmetry5}
\end{figure}

The dependence of the parameter $\zeta$, eq.~(\ref{mixing_sumrule_50}), 
which governs the mixing effect of the pion formfactor (\ref {mixing_30}) 
via (\ref{mixing_sumrule_40}), on the asymmetry parameter $\alpha_{n p}$ 
is seen in Fig.~\ref{fig: Asymmetry5} (a).   
For strong asymmetry one obtains a remarkable increase of $\zeta$, roughly 
linear with $\alpha_{n p}$.  
We note that the dashed curve of Fig.~\ref{fig: Asymmetry5} (a) is in 
good agreement with \cite{Mixing5} where an asymmetry dependence 
$\zeta = \zeta^{(0)} + \zeta^{(1)} \, \alpha_{n p}\, n /(0.2\, n_0)$  
with $\zeta^{(0)} = 1.1 \times 10^{-3}$ and $\zeta^{(1)} = 1.5 \times 10^{-3}$ 
has been reported, while our findings correspond to $\zeta^{(0)} = 
1.05 \times 10^{-3}$ and $\zeta^{(1)} = 1.9 \times 10^{-3}$.

Altogether, without accounting for the mass
shifts, an amplification of the mixing effect in the pion formfactor is
obtained (see dashed curve in Fig.~\ref{fig: Asymmetry5} (b)).
In contrast, when accounting for the individual mass parameter shifts the 
mixing effect is washed out (see solid curve of 
Fig.~\ref{fig: Asymmetry5} (b)). 

Finally, it is expedient to summarize the differences between the analysis 
presented here and Ref. \cite{Mixing5},
which are, so far, the only investigations where the QCD sum rule approach 
has been applied to the mixing effect at finite density.  
Besides the usage of a complete OPE up to mass-dimension-6 twist-2 for the  
mixing effect and a selfconsistent Borel analysis for all unknowns at finite   
density in our work, the improvements are the following:  
First, we have implemented the individual mass parameter 
shifts of the vector mesons in a consistent way and have studied their impact 
on the mixing 
effect. A second difference consists in taking into account 
the $\phi$ meson on the hadronic side, which is necessary  
due to large cancellations between $\rho$ and $\omega$ meson contributions 
(this has been pointed out for vacuum in \cite{Maltman}).
Thirdly, we have investigated the relevance of $\zeta$ by considering  
the influence of the mixing on pion  
formfactor and di-electron production rate.

\section{Summary}

In summary, we have investigated the mass parameter splitting and the mixing 
of $\rho$ and $\omega$ mesons in nuclear matter within the QCD sum rule 
approach, starting from a complete OPE of the current-current correlator 
up to mass dimension-6 twist-4 and up to the first order in the 
coupling constant. 
Special attention is devoted to the impact of the poorly known scalar 
4-quark condensates. 
We have found a strong $\rho - \omega$ mass parameter splitting. 
The scalar flavor mixing condensate has been evaluated at finite density using 
quite general assumptions. It turns out that this condensate, 
while responsible for the $\rho - \omega$ mass parameter splitting in vacuum, 
plays a subdominant role in matter.
Instead, the individual mass parameter splitting of $\rho$ and $\omega$ mesons 
is mainly governed by the Landau damping terms. The scalar 4-quark condensates 
have a strong impact on the individual strengths of the mass parameter shifts, 
while the amount of the splitting is fairly insensitive to these condensates. 

We emphasize that the mass parameters are weighted moments of the spectral 
functions. A mass parameter shift in medium does not 
necessarily mean a simple shift of the peak position of a spectral function, 
rather additional strength may occur at lower or higher energies causing a 
shift of the weighted moment. The presently employed form of the QCD sum rule 
approach is not sensitive to such details. Only a detailed modelling of the
hadronic in-medium spectral function with parametric dependences allows for 
more concise statements \cite{BrownRho}.

Another physical effect investigated concerns the 
$\rho - \omega$ mixing at finite density and the impact of the 
$\rho - \omega$ mass parameter splitting.  
Starting with the vacuum  
we find an excellent agreement with experimental data recently obtained.
In medium, the nondiagonal selfenergy $\delta_{\rho \omega} (\overline{m}, n)$, which 
drives the mixing effect, is only weakly amplified in isospin symmetric 
nuclear matter.  
The mixing parameter $\zeta$, however, is remarkably enlarged for 
strongly isospin asymmetric nuclear matter, such as in uranium nuclei 
with $\alpha_{n p} = 0.2$.
Therefore, not taking into account the individual mass shifts of the 
$\rho$ and $\omega$ meson would indeed result in an in-medium amplification  
of the mixing effect.
In contrast, if one takes into account the strong mass parameter splitting of 
$\rho$ and $\omega$ mesons as a pronounced splitting 
of the corresponding peaks then the mixing effect in the pion formfactor 
as well as in the di-electron production rate disappears in medium, both 
for isospin symmetric and isospin asymmetric nuclear matter. 
Upcoming measurements at HADES can deliver valuable information 
on these issues. 
\section*{Acknowledgements}
This work is supported 
by BMBF 06DR121 and GSI-FE.
\newpage

\appendix

\section{Scalar flavor-unmixing 4-quark condensates}

In lines (\ref{ope_4}) and (\ref{ope_7}) one recognizes two  
different types of scalar flavor-unmixing 4-quark condensates ($q=u,d$)
\bea
M_A^{q q} = \langle \Omega| \overline{q} \gamma_{\mu} \gamma_5 \lambda^a q\;
\overline{q} \gamma^{\mu} \gamma_5 \lambda^a q |\Omega\rangle
\eea
and 
\bea
M_V^{q q} = \langle \Omega| \overline{q} \gamma_{\mu} \lambda^a q\;
\overline{q} \gamma^{\mu} \lambda^a q |\Omega\rangle\;.
\eea
Previous studies employed a factorization for the scalar flavor-unmixing 
4-quark condensates \cite{condensates}. 
We go beyond such approximation, 
$M_A^{q q} = \frac{16}{9} \kappa \langle \Omega| \overline{q} q |\Omega
\rangle^2$, 
pointing out that $\kappa$ is uncertain and might even have a density 
dependence. In the spirit of the linear density approximation 
eq.~(\ref{eq_40}), a Taylor expansion results in  
\bea
M_A^{q q} = \frac{16}{9} \langle \overline{q} q \rangle^2_0\;\kappa_0^ {(1)} 
\left[1 + \frac{\kappa_N^{(1)}}{\kappa_0^{(1)}} \frac{\sigma_N\; n}
{m_q\, \langle \overline{q} q \rangle_0}\right]\;.
\label{A_10}
\eea
The first term, i.e. 
$\frac{16}{9} \langle \overline{q} q \rangle^2_0 \;\kappa_0^ {(1)}$, 
is merely an expression for 
$\langle \overline{q} \gamma_{\mu} \gamma_5 \lambda^a q\;
\overline{q} \gamma^{\mu} \gamma_5 \lambda^a q \rangle_0$. 
The second term, proportional to $\kappa_N^{(1)}$, parameterizes 
the poorly known 4-quark condensate in the nucleon  
$\langle N(\mak)| \overline{u} \gamma_{\mu} \gamma_5 \lambda^a u\;
\overline{u} \gamma^{\mu} \gamma_5 \lambda^a u |N(\mak)\rangle$.
Similarly, for the other 4-quark condensate we obtain
\bea
M_V^{q q} = - \frac{16}{9} 
\langle \overline{q} q \rangle^2_0 \;\kappa_0^ {(2)}
\left[1 + \frac{\kappa_N^{(2)}}{\kappa_0^{(2)}} \frac{\sigma_N\; n}
{m_q\,\langle \overline{q} q \rangle_0}\right]\;.
\label{A_15}
\eea
Accumulating all flavor-unmixing four-quark condensates, 
with the right weight given from the OPE, one obtains 
in linear density approximation finally
\bea
- \frac{1}{2} M_A^{u u} - \frac{1}{2} M_A^{d d} - \frac{1}{9} M_V^{u u}
- \frac{1}{9} M_V^{d d} 
 = - \frac{7}{18} \frac{16}{9}
\langle \overline{q} q \rangle^2_0 \;\kappa_0
\left[1 + \frac{\kappa_N}{\kappa_0} \frac{\sigma_N\; n}
{m_q\,\langle \overline{q} q \rangle_0}\right]\;.
\label{A_20}
\eea
Note that $\kappa_N = \kappa_0$ is conform to the large-$N_c$ limit 
\cite{Leupold_Large_Nc}. 
Since we are interested in medium effects, we adjust the value of 
$\kappa_0$ to the vacuum masses, yielding $\kappa_0 = 3$ both for $\rho$ and 
$\omega$, and study the impact of the unknown parameter $\kappa_N$. 
As stressed in \cite{zschocke1,zschocke3}, only a comparison with experimental 
data can pin down $\kappa_N$.

For treating the $\rho - \omega$ mixing we also need 
\bea
N_A^{q q} = \langle \Omega| \overline{q} \gamma_{\mu} \gamma_5 q\;
\overline{q} \gamma^{\mu} \gamma_5 q |\Omega\rangle
\label{A_25}
\eea
and 
\bea
N_V^{q q} = \langle \Omega| \overline{q} \gamma_{\mu} q\;
\overline{q} \gamma^{\mu} q |\Omega\rangle\;.
\label{A_30}
\eea
With the same steps as above we arrive at
\bea
- 4 N_A^{u u} + N_A^{d d} - \frac{8}{9} N_V^{u u} + \frac{2}{9} N_V^{d d} 
= - \frac{7}{9} \langle \overline{q} q \rangle^2_0\; \kappa_0
\left[1 + \frac{\kappa_N}{\kappa_0} \frac{\sigma_N\; n}
{m_q\,\langle \overline{q} q \rangle_0}\right]\;.
\label{A_35}
\eea

\newpage

\section{Scalar flavor-mixing 4-quark condensates}

Now we estimate the two scalar flavor-mixing condensates 
in lines (\ref{ope_5}) and (\ref{ope_8}) at finite density. 
(For the vacuum such an estimate is given in \cite{litD1}.) 
Let us first consider the condensate in line (\ref{ope_5}).  
To evaluate such a condensate we insert a complete set of QCD eigenstates 
after a Fierz transformation
\bea
M_A^{u d} &=& \langle \Omega | \overline{u} \gamma_{\mu}
\gamma_5 \lambda^{a} u \; \overline{d} \gamma^{\mu}
\gamma_5 \lambda^{a} d | \Omega \rangle\nonumber\\
&=&
\sum \limits_n \langle \Omega | \overline{u}_i^{\alpha} d_l^{\delta}
| n\rangle \langle n| \overline{d}_k^{\gamma} u_j^{\beta} |\Omega \rangle
\; (\lambda^a)_{i j}\; (\lambda^a)_{k l}\nonumber\\
&&\hspace{-1.0cm}\times\left(
\delta^{\alpha \delta} \delta^{\gamma \beta}
+ \frac{1}{2} (\gamma_{\mu})^{\alpha \delta} (\gamma^{\mu})^{\gamma \beta}
+ \frac{1}{2} (\gamma_{\mu} \gamma_5)^{\alpha \delta}
(\gamma^{\mu} \gamma_5)^{\gamma \beta}
- (\gamma_5)^{\alpha \delta} (\gamma_5)^{\gamma \beta}
\right)\;,
\label{FM_5}
\eea
and approximate the sum by
\bea
\sum\limits_n | n \rangle \langle n| \approx
|\Omega\rangle \langle \Omega| + |\Omega^{*}\rangle \langle \Omega^{*}|
+ \sum \limits_{b=1}^{3}
\int \frac{d^3 p}{(2 \pi)^3} \;\frac{1}{2 E_p}\;
|\Omega\; \pi^b (p) \rangle \langle \pi^b (p) \;\Omega|\;,
\label{FM_10}
\eea
where $|\Omega\rangle$ is the ground state of matter, 
$|\Omega^{*}\rangle$ denotes low-lying excitations 
(e.g. particle-hole excitations), 
and $|\Omega\,\pi^b\rangle$ means ground state plus pion with isospin index 
$b$  
(other states with 
mesons heavier than pions are suppressed by their larger masses).  
The matrix elements
$\langle\Omega| \overline{u}_i^{\alpha} d_l^{\delta} |\Omega\rangle$
and
$\langle\Omega| \overline{u}_i^{\alpha} d_l^{\delta} |\Omega^{*}\rangle$
vanish due to quark flavor conservation yielding 
\bea
M_A^{u d} &=&
\sum \limits_{b=1}^3 \int \frac{d^3 p}{(2 \pi)^3}
\frac{1}{2 E_p} \langle \Omega | \overline{u}_i^{\alpha} d_l^{\delta}
| \Omega \pi^b (p) \rangle \langle \Omega \pi^b (p)|
\overline{d}_k^{\gamma} u_j^{\beta} |\Omega \rangle\;
(\lambda^a)_{i j} \;(\lambda^a)_{k l} \nonumber\\
&&\hspace{-1.0cm}\times\left(
\delta^{\alpha \delta} \delta^{\gamma \beta}
+ \frac{1}{2} (\gamma_{\mu})^{\alpha \delta} (\gamma^{\mu})^{\gamma \beta}
+ \frac{1}{2} (\gamma_{\mu} \gamma_5)^{\alpha \delta}
(\gamma^{\mu} \gamma_5)^{\gamma \beta}
- (\gamma_5)^{\alpha \delta} (\gamma_5)^{\gamma \beta}
\right)\;.
\label{FM_15}
\eea
The soft pion theorem \cite{productionrate_1,softpion} allows to calculate the 
needed terms in linear density approximation as 
\bea
\langle \Omega | \overline{u}_i^{\alpha} d_l^{\delta} |\Omega\;\pi^1(0)
\rangle &=&
\frac{i}{f_{\pi}}\frac{1}{12} \delta_{i l}
(\gamma_5)^{\alpha \delta}  \langle \Omega |
\overline{q} q |\Omega\rangle \;,\\
\langle \Omega | \overline{u}_i^{\alpha} d_l^{\delta} |\Omega\;\pi^2(0)
\rangle &=&
- \frac{1}{f_{\pi}}\frac{1}{12} \delta_{i l}
(\gamma_5)^{\alpha \delta}  \langle \Omega |
\overline{q} q |\Omega\rangle 
\label{FM_20}
\eea
and $\langle \Omega | \overline{u}_i^{\alpha} d_l^{\delta} |\Omega\;\pi^3(0)
\rangle = 0$.
Inserting these matrix elements into (\ref{FM_15}) results in  
\bea
M_A^{u d} = 
\frac{4}{9 \pi^2} \frac{1}{f_{\pi}^2} \;\langle \Omega|
\overline{q} q | \Omega\rangle^2\;
Q_0^2 
\label{FM_25}
\eea
with the cutoff $Q_0$ coming from the momentum integral.
With eq.~(\ref{eq_40}) and $\langle N(\mak)| \overline{q} q | N(\mak) \rangle 
= \frac{M_N \,\sigma_N}{m_q}$ one gets the final result
\bea
M_A^{u d} = \frac{4}{9 \pi^2} \frac{Q_0^2}{f_{\pi}^2} 
\langle \overline{q} q \rangle^2_0 \left[ 1 + \frac{\sigma_N \, n}
{m_q \, \langle \overline{q} q \rangle_0} \right]\;.
\label{FM_30}
\eea
Using the same technique for the matrix element in line (\ref{ope_8}) one finds 
\bea
M_V^{u d} = \langle \Omega | \overline{u} \gamma_{\mu}
\lambda^{a} u \; \overline{d} \gamma^{\mu}
\lambda^{a} d | \Omega \rangle = - M_A^{u d}\;.
\label{FM_35}
\eea
$Q_0$ is adjusted to the vacuum $\rho - \omega$ mass splitting. 
Using $Q_0 = 150$ MeV we get the right experimental vacuum masses, i.e. 
$m_{\rho} (0) = 771$ MeV and $m_{\omega} (0) = 782$ MeV \cite{pdb} 
for our chosen parameters, 
$\langle \overline{q} q \rangle_0 = (- 0.245\; {\rm GeV})^3$, 
$\langle\alpha_s/\pi \;G^2\rangle_0 = (0.33\; {\rm GeV})^4 $, 
$\alpha_s=0.38$, $m_u=4$ MeV, 
$m_d=7$ MeV, $M_N^0 = 770$ MeV, $\sigma_N=45$ MeV. 
Furthermore, we take the known vacuum values of $M_N, f_{\pi}$ and $m_{\pi}$. 


\section{Twist-2 condensates}

The quark twist-2 condensates appear in lines (\ref{ope_9}) and 
(\ref{ope_11}), respectively, while the gluonic twist-2 condensates appear in 
lines (\ref{ope_10}) and (\ref{ope_12}), respectively. 
The operator $\hat {\rm S} \hat {\rm T}$ creates a
symmetric and traceless expression with respect to the Lorentz indices, i.e.,  
for spin-2 
$\hat {\rm S} \hat {\rm T} \left({\cal O}_{\alpha \beta}\right) =
\frac{1}{2!} \left({\cal O}_{\alpha \beta} + {\cal O}_{\beta \alpha} \right)
- \frac{1}{4} \, g_{\alpha \beta} \,{\cal O}_{\gamma}^{\gamma}$ 
and analogously for spin-4 condensates. 
These condensates vanish in vacuum and therefore, according to the low-density 
approximation (\ref{eq_40}), we need only the nucleon matrix elements which 
can generally be written as \cite{condensates} 
\bea
\langle N (\mak)| \hat{\rm S} \hat{\rm T} \overline{q} \gamma_{\mu} D_{\nu} 
q | N (\mak)\rangle &=& - i S_{\mu \nu} \; A_2^q(\mu^2) \;, 
\label{eq_134}\\
\nonumber\\
\langle N (\mak)| \hat{\rm S} \hat{\rm T} G_{\mu}^{\;\;\alpha}
G_{\alpha \nu} | N (\mak)\rangle &=& S_{\mu \nu} \; A_{2}^{G} (\mu^2)
\label{eq_135}
\eea
for spin-2 operators and
\bea
\langle N (\mak)| \hat{\rm S} \hat{\rm T} \overline{q} \gamma_{\mu} D_{\nu} 
D_{\lambda} D_{\sigma} q | N (\mak)\rangle &=& i 
S_{\mu \nu \lambda \sigma} \; A_4^q(\mu^2) \;,
\label{eq_136}\\
\nonumber\\
\langle N (\mak)| \hat{\rm S} \hat{\rm T} G^{\;\;\rho}_{\mu} 
D_{\nu} D_{\lambda} G_{\rho \sigma} | N (\mak)\rangle &=&
- S_{\mu \nu \lambda \sigma} \; A_{4}^G (\mu^2)
\label{eq_140}
\eea
for spin-4 operators, respectively. The Lorentz structures are defined as
\bea
S_{\mu \nu} &=& k_{\mu} k_{\nu} - \frac{1}{4} k^2\; g_{\mu \nu}\;, 
\label{Lorentz1}\\
S_{\mu \nu \lambda \sigma} &=& \Bigg[ k_{\mu} k_{\nu} k_{\lambda} k_{\sigma}
+ \frac{k^4}{48} \left(g_{\mu \nu} g_{\lambda \sigma} + g_{\mu \lambda}
g_{\nu \sigma}
+ g_{\mu \sigma} g_{\nu \lambda}\right)
\nonumber\\
\nonumber\\
&&\hspace{0.0cm} -\frac{k^2}{8} \left(k_{\mu} k_{\nu} g_{\lambda \sigma} +
k_{\mu} k_{\lambda} g_{\nu \sigma} + k_{\mu} k_{\sigma} g_{\lambda \nu}
+ k_{\nu} k_{\lambda} g_{\mu \sigma} + k_{\nu}k_{\sigma} g_{\mu \lambda}
+ k_{\lambda} k_{\sigma} g_{\mu \nu} \right)
\Bigg]\;.
\label{Lorentz2}
\eea
The reduced matrix elements of quark twist-2 condensates are defined as 

$A_i^q (\mu^2) = 2 \int\limits_0^1 d x \; x^{i-1} \left[ q_N (x, \mu^2) + 
(-1)^{i} \;\overline{q}_N (x, \mu^2) \right]$, where $q_N (x, \mu^2)$ and 
$\overline{q}_N (x, \mu^2)$ are the quark and antiquark distribution 
function inside the nucleon. We take  
$A_2^{(u+d)} (1 {\rm GeV}^2) = 1.02$ and $A_4^{(u+d)} (1 {\rm GeV}^2) = 0.12$ 
\cite{lit12}, respectively. The reduced matrix elements of gluon twist-2 
condensates are defined by 
$ A_{i}^G (\mu^2) = 2 \int\limits_0^1 d x \; x^{i-1} G_{N} (x,\mu^2)$, 
with $G_{N}(x,\mu^2)$ as gluon distribution function inside the nucleon
at the scale $\mu^2$. We use   
$A_{2}^G (1 {\rm GeV}^2) = 0.83$ and $A_{4}^G (1 {\rm GeV}^2) = 0.04$ 
\cite{Leupold1}, respectively.

\section{Twist-4 condensates}

Twist-4 condensates appear in the lines 
(\ref{ope_18}, \ref{ope_17}, \ref{ope_15}) and (\ref{ope_19}).
All twist-4 operators vanish in vacuum and therefore, according 
to the low-density approximation (\ref{eq_40}), one needs only the 
nucleon matrix elements. 
The nucleon matrix elements of symmetric and traceless
twist-4 operators can be decomposed as \cite{DIS}
\bea
\hspace{0.0cm}\langle N(\mak) |i g_s\; \hat{\rm S} \hat{\rm T}
\left(\overline{u} \left[ D_{\mu} , \tilde{G}_{\nu \alpha} 
\right]_{+} \gamma^{\alpha} \gamma_5 u \right) | N(\mak) \rangle 
&=& \frac{1}{2} \, S_{\mu \nu} 
\left( K^g_u + K^g_d\right)\;,
\label{twist4_1}
\\
\nonumber\\
\hspace{0.0cm}\langle N(\mak) |g_s^2 \;\hat{\rm S} \hat{\rm T} \left( 
 \overline{u} \gamma_{\mu} \gamma_5 \lambda^a u 
\overline{u} \gamma_{\nu} \gamma_5 \lambda^a u \right)| N(\mak) \rangle 
&=& 2 \, S_{\mu \nu}
\left( K^1_u + K^1_d - K^1_{u d} \right)\;,
\label{twist4_2}
\\
\nonumber\\
\hspace{0.0cm}\langle N(\mak) | g_s^2\; \hat{\rm S} \hat{\rm T} \left(
\overline{u} \gamma_{\mu} \gamma_5 \lambda^a u
\overline{d} \gamma_{\nu} \gamma_5 \lambda^a d \right)| N(\mak) \rangle
&=& 2 \, S_{\mu \nu} 
\left(K^1_{u d} \right)\;,
\label{twist4_3}
\\
\nonumber\\
\hspace{0.0cm}\langle N(\mak) | g_s^2 \hat{\rm S} \hat{\rm T} \left( 
\overline{u} \gamma_{\mu} \lambda^a u
\left( \overline{u} \gamma_{\nu} \lambda^a u + 
\overline{d} \gamma_{\nu} \lambda^a d \right) \right)| N(\mak) \rangle
&=&  2 \,S_{\mu \nu} \,
\left( K^2_u + K^2_d\right) 
\label{twist4_4}
\eea
with $S_{\mu \nu}$ defined in eq.~(\ref{Lorentz1}).
The other twist-4 condensates, where $u$ and $d$ are interchanged, are 
equal to the given ones due to the assumed flavor symmetry. 
As pointed out in \cite{DIS} the coefficients $K_{u,d,ud}^{1,2}$ are related 
to the nucleon forward scattering amplitude of the electromagnetic current.
We take the following parameter set:
$K^1_u=-0.112$ GeV$^2$, $K^2_u=0.110$ GeV$^2$, $K^g_u=-0.300$ GeV$^2$,
$K^1_{ud}=-0.084$ GeV$^2$ as default. For the $d$ quark we use $K_d^{1,2,g} =
\beta K_u^{1,2,g}$ with $\beta=0.476$ from \cite{DIS}.

We remark that the parameters $K_{u,d,ud}^{1,2,g}$ should be taken at a 
hadronic scale of $\mu=1$ GeV.
Unfortunately, twist-4 condensates are poorly known and even available
only at a scale of $\mu = 2.25$ GeV. To evolve these 
parameters down to $\mu=1$ GeV would require
the knowledge of anomalous dimensions which are 
not available. Here we use the above condensates, expressed by
$K_u^1$, $K_u^2$, $K_u^g$ and $K_{u d}^1$,
to demonstrate that accounting these condensates has little influence on the 
mass splitting and individual mass shifts of $\rho$ and $\omega$ mesons. 

For the twist-4 operator in line (\ref{ope_20}) we use the estimate 
\cite{twist_quark}:
\bea
\langle N(\mak) | m_q \; \overline{q} D_{\mu} D_{\nu} q | N(\mak) \rangle
\simeq - P_{\mu}^q P_{\nu}^q \langle N(\mak) | m_q \; \overline{q} q 
| N(\mak) \rangle\;, 
\label{twist4_6}
\eea
where $P_{\mu}^q$ is the average momentum carried by the quark $q$ 
inside the nucleon. Taking $P_{\mu}^q \sim k_{\mu}/6$ \cite{twist_quark} 
($k_{\mu}$ is the momentum of nucleon) and making the operator symmetric and 
traceless we get
\bea
\langle N(\mak) | 
\hat{S} \hat{T} \;m_q \;\overline{q} D_{\mu} D_{\nu} q | N(\mak) \rangle
\simeq - S_{\mu \nu} \frac{1}{36} m_q\;\langle N(\mak) | \overline{q} q 
| N(\mak)\rangle\;.
\label{twist4_7}
\eea

\section{Parameters for $\rho - \omega$ mixing}

For the $\rho - \omega$ mixing we have to distinguish between proton 
and neutron matrix elements. In particular we need the following 
twist-2 and twist-4 condensates    
\bea 
\langle p(\mak) | \hat{\rm S} \hat{\rm T}\; \overline{q} \gamma_{\mu} 
D_{\nu} q| p(\mak) \rangle 
= - i S_{\mu \nu} A_2^{q,p} \;, 
\label{AppendixE_5}\\
\nonumber\\
\langle p(\mak) | \hat{\rm S} \hat{\rm T}\;\overline{u} \gamma_{\mu} 
D_{\nu} D_{\lambda} D_{\sigma} u | p(\mak) \rangle =  
i S_{\mu \nu \lambda \sigma} A_4^{q,p}
\label{AppendixE_10}
\eea
with the proton state $| p(\mak) \rangle$ and analog expressions for the 
neutron. The reduced matrix elements $A_i^{q,p} (\mu^2)$  
are defined as 
$A_i^{q,p} (\mu^2) = 2 \int\limits_0^1 dx\;x^{i-1} [q_p (x, \mu^2) + 
(-1)^{i} \; \overline{q}_p (x, \mu^2)]$, where $q_p (x, \mu^2)$ and 
$\overline{q}_p (x, \mu^2)$ are the quark and antiquark 
distribution functions inside the proton. We use the following parameters:  
$A_2^{u, p} (1 {\rm GeV}^2) = 0.67$, $A_2^{d, p} (1 {\rm GeV}^2) = 0.35$, 
$A_4^{u, p} (1 {\rm GeV}^2) = 0.091$, $A_4^{d, p} (1 {\rm GeV}^2) = 0.029$; 
$A_2^{u, n} = A_2^{d, p}$ and $A_2^{d, n} = A_2^{u, p}$, 
which follow from $u_n (x, \mu^2) = d_p (x, \mu^2)$ and 
$u_p (x, \mu^2) = d_n (x, \mu^2)$, respectively \cite{Peshkin_Schroeder}.  

Another needed matrix element is \cite{Asymmetry5}
\bea
\langle p | \overline{u} u - \overline{d} d | p \rangle =
2 \;M_N\;\frac{m_{\Xi} - m_{\Sigma}}{m_s} = 1.3 \;{\rm GeV}\;.
\label{asymmetric15}
\eea
 
The isospin symmetry breaking parameter for the quark condensate is 
$\gamma=-0.008$ (cf. \cite{Mixing15}), and the mass parameters of higher 
resonances are $m_{\rho'} = 1.465$ GeV and $m_{\omega'} = 1.649$ GeV 
\cite{pdb}, respectively. 
For the coupling constants we take the values 
$g_{\rho \pi \pi} = 6.0$, $g_{\rho \gamma} = 5.2$ and 
$g_{\omega \gamma} = 3 g_{\rho \gamma}$ \cite{Connell1}.


\end{document}